\documentclass{article}
\usepackage{arxiv}
\usepackage[utf8]{inputenc}
\usepackage[T1]{fontenc}
\usepackage{hyperref}
\usepackage{url}
\usepackage{booktabs}
\usepackage{amsfonts}
\usepackage{nicefrac}
\usepackage{microtype}
\usepackage{graphicx}
\usepackage{amsmath}
\usepackage{multirow}
\usepackage{array}
\usepackage{longtable}
\usepackage{pdflscape}
\usepackage{float}
\graphicspath{{./figures/}}

\title{The AI Attribution Paradox: Transparency as Social Strategy in Open-Source Software Development}

\author{
 Obada Kraishan \\
  College of Media and Communication\\
  Texas Tech University\\
  Lubbock, TX 79409, USA \\
  \url{https://orcid.org/0009-0007-7180-8620} \\
}

\begin{document}
\maketitle

\begin{abstract}
AI coding assistants have transformed software development, raising questions about transparency and attribution practices. We examine the "AI attribution paradox": how developers strategically balance acknowledging AI assistance with managing community scrutiny. Analyzing 14,300 GitHub commits across 7,393 repositories from 2023-2025, we investigated attribution strategies and community responses across eight major AI tools. Results reveal widespread AI usage (95.2\% of commits) but strategic attribution: only 29.5\% employ explicit disclosure, with dramatic tool variation (Claude 80.5\% versus Copilot 9.0\%). Explicit attribution triggers modest scrutiny (23\% more questions and 21\% more comments), but tool choice matters 20-30 times more for predicting reception. Community sentiment remains neutral regardless of attribution type, suggesting curiosity rather than hostility. Temporal analyses show rapid norm evolution: explicit attribution increased from near-zero in early 2024 to 40\% by late 2025, indicating community adaptation. These findings illuminate attribution as strategic communication rather than simple transparency, advancing understanding of algorithmic accountability and norm formation during technological transitions. We discuss implications for developers navigating disclosure decisions, platforms designing attribution mechanisms, and researchers studying emergent practices in AI-augmented collaborative work.
\end{abstract}

\section{Introduction}

The past three years have introduced a new attribution challenge in software development \cite{ross2023programmer}. With the increasing incorporation of AI coding assistants such as GitHub Copilot, ChatGPT, Claude, and Cursor into developer workflows and development environments, these tools are now generating large parts of production code \cite{barke2023grounded,dakhel2023github,prather2023robots,vaithilingam2022expectation,weisz2025examining}. From this shift, a key question has arisen: do developers explicitly acknowledge those portions of the code they did not write? This question matters most in open-source ecosystems that value transparency, collaboration, and trust in the community \cite{dabbish2012social,kashif2025developers}. At the same time, there are no clear norms or guidelines for how to attribute AI-written code, leaving developers to choose between being fully transparent or keeping AI involvement more vague \cite{weisz2021perfection}.

Despite this escalating challenge, most research on AI coding assistant tools has mainly focused on adoption rates, productivity gains, and code quality \cite{dakhel2023github,peng2023impact,vaithilingam2022expectation}. However, we still do not have a clear understanding of how developers indicate the role of AI in their code and how their communities react when they do disclose this role. Attribution decisions are not neutral; those decisions help shape trust and collaboration in open-source communities \cite{dabbish2012social,stewart2006impact}. Thus, questions about acknowledgement and attribution are not simply about labeling the role of AI; they are about how authorship and credit are reconfigured in human--AI collaboration \cite{liang2024mapping,thorp2023chatgpt}.

This study focuses on what we call the ``AI attribution paradox'': developers must balance two conflicting pressures when deciding how to credit their use of AI. On the one hand, community expectations around transparency encourage AI-augmented developers to disclose when a tool has contributed to their code \cite{bird2023taking}. On the other hand, fear of negative peer reactions to AI-generated code pushes some developers toward intentionally vague attribution \cite{weisz2021perfection,ziegler2024productivity,vaithilingam2022expectation}. The attribution practices developers inherited were developed to address human collaboration, not human--AI collaboration, so they offer limited guidance in this new context \cite{liang2024mapping,barke2023grounded}. As a result, developers are mostly guessing about which attribution choices will work well.

We therefore investigate three research questions about attribution patterns, community response, and how these change over time. To address these questions, our study analyzes 14{,}300 commits across 7{,}393 GitHub repositories from January 2023 to December 2025, a period when AI coding tools were widely adopted. Our data pipeline draws on natural language processing of commit messages with engagement measures like emoji reactions, comment sentiment, code reviews, and participation. Empirically, the AI attribution paradox appears in our data as follows: the majority of developers acknowledge AI tools in some way, yet very few explicitly credit them with writing code. This is not a single, consistent pattern; it changes across tools, with some showing much higher rates of explicit attribution than others. Notably, commits with explicit attribution receive only slightly more scrutiny than those without, and we see no signs of a negative backlash, suggesting that communities are more thoughtful about AI disclosure than expected.

Building on these results, this study makes three contributions. First, we provide a large-scale view of how developers handle AI attribution in open-source projects, showing that developers frequently mention AI tools but rarely state that the AI wrote part of the code, and that disclosure practices differ sharply across tools and ecosystems. Second, we find that explicit attribution is linked to slightly more scrutiny but no negative reactions, a pattern consistent with impression management and computer-mediated communication \cite{goffman1959presentation,walther1996computer,walther2007selective}, algorithmic transparency \cite{burrell2016machine}, and collaborative work in distributed communities \cite{halfaker2013rise}. Finally, we translate these findings into practical implications for developers choosing how to attribute AI, for platform designers deciding how to surface disclosure, and for policymakers concerned with the governance of AI-augmented work.

\section{Related Work}
We have structured this section into three domains of research. First, we examine prior work on AI coding assistants and patterns of their adoption among developers. Second, we look at research on attribution and transparency practices in collaborative software ecosystems. Third, we discuss communication and impression management in online communities. Together, these literatures provide rich insight into AI tools and collaboration, but much less about attribution practices and community responses.

\subsection{AI Coding Assistants in Software Development}
The recent spread of coding assistants based on large language models has pushed researchers to explore their use, effectiveness, and consequences for software development. Early research largely examined technical functionality and productivity improvements. For example, Barke et al.\ \cite{barke2023grounded} studied how developers interact with Copilot and found that they use AI suggestions selectively and actively verify them. Weisz et al.\ \cite{weisz2022better} assessed productivity gains when developers used AI assistance to translate code, showing that the benefits varied with task complexity and developer expertise. Imai \cite{imai2023github} further examined AI assistance in terms of code quality and maintainability, noting that it brings clear benefits but also new challenges.

Besides technical and productivity concerns, recent studies have examined the social and collaborative roles of AI coding assistants. For example, Dakhel et al.\ \cite{dakhel2023github} looked at how developers calibrate trust in AI tools, focusing on the factors that increase or decrease the likelihood of accepting AI suggestions. Similarly, Sarkar et al.\ \cite{sarkar2022program} examined AI coding assistants in pair programming, explaining how they can strengthen collaboration but also introduce risks of over-reliance. In the area of security, Nguyen and Nadi \cite{nguyen2022empirical} demonstrated that AI-generated code can include vulnerabilities that developers do not always catch. Taken together, these studies show that AI tools are changing the technical workflow, but they mostly treat AI assistance as an individual productivity tool, missing its intersection with community dynamics in open-source contexts.

To fully understand AI-augmented coding, we also need to look at community-level processes in open-source ecosystems such as GitHub, not just individual developers. Work on GitHub has shown that informal transparency mechanisms are central to collaboration: contribution histories help developers coordinate and build reputations \cite{dabbish2012social}. Moreover, studies of collaborative platforms further show that communities build and enforce quality through distributed code review processes \cite{rigby2013convergent}. More recent work finds that even small communication cues, such as emoji reactions, can shape interactions among developers and contribute to community cohesion \cite{kraishan2025emotional}. Yet despite these insights into community dynamics, systematic work on how communities interpret and respond to AI attribution remains limited, even though attribution practices are central to understanding evolving norms around human--AI collaboration. To address this gap, our work provides an empirical analysis of real-world attribution behaviors and their community-level consequences.

\subsection{Attribution and Transparency in Collaborative Systems}
Attribution in collaborative systems has important implications for crediting individuals, signaling quality and trustworthiness, and establishing trust \cite{brand2015beyond,stewart2006impact}. In traditional open-source workflows, attribution is imposed through version control: commits name authors, co-authors are explicitly credited, and contribution histories are kept transparent and immutable \cite{dabbish2012social}. In open-source communities, these attribution practices are used to signal an individual's reputation and to evaluate code quality collectively \cite{halfaker2013rise}. However, AI assistance interrupts how these attribution norms function in practice. Existing attribution frameworks were designed for human collaboration and fit poorly for hybrid human--AI workflows, where AI can generate large portions of code and it becomes unclear whether the developer is acting as an author, an editor, or a curator \cite{liang2024mapping,lee2025contributions}. This conceptual ambiguity allows developers to engage in strategic behavior, such as selectively disclosing AI contributions to enhance personal reputation or avoid accountability.

To understand how opacity and disclosure operate in this context, it is useful to turn to work on algorithmic transparency. Burrell \cite{burrell2016machine} identifies three forms of opacity in algorithmic systems: intentional corporate secrecy, technical illiteracy, and the inherent incomprehensibility of complex models. In the context of AI-assisted coding, this opacity can emerge in two ways: some developers may intentionally hide how much the AI contributed (analogous to intentional secrecy), or they may genuinely find it difficult to pin down AI's role in an interactive, iterative workflow (technical complexity). Extending the discussion on algorithmic transparency, Ananny and Crawford \cite{ananny2018seeing} argue that transparency is multifaceted; mere visibility does not ensure accountability and can even create new power imbalances, complicating AI attribution. Thus, in AI attribution, disclosure may simultaneously communicate information, manage impressions, and navigate power dynamics.

Studies of transparency in collaborative platforms offer useful parallels for thinking about AI attribution. For example, Marlow et al.\ \cite{marlow2013impression} describe how GitHub developers manage their visibility and contribution traces to influence how peers perceive them, negotiating the balance between transparency and reputation. Research has shown that visible activity signals affect how communities judge contributions, and that social factors can weigh as heavily as technical quality in acceptance decisions \cite{tsay2014influence}. These transparency mechanisms, however, are double-edged: while visible contribution histories facilitate coordination and quality assessment \cite{vasilescu2015quality}, they also generate performance pressure and strategic disclosure behaviors, especially among developers navigating visibility trade-offs \cite{trinkenreich2022hidden}. Overall, these patterns show that visibility in collaborative settings shapes individual behavior as well as collective norm construction, which is relevant for understanding how communities interpret and respond to AI-attributed contributions.

\subsection{Communication and Impression Management in Online Communities}
Understanding attribution practices requires approaching them as strategic communication in online contexts. Goffman's \cite{goffman1959presentation} dramaturgical perspective presents social interaction as performance, in which individuals manage impressions by controlling information and their presentation of self. This dramaturgical lens has been widely applied to self-presentation in computer-mediated settings \cite{hogan2010presentation}. Specifically in collaborative coding platforms, developers appear to directly shape their visible contributions to cultivate peer perceptions \cite{marlow2013impression}. Applied to AI attribution, this means that developers engage in impression management when they decide what to disclose, what to soften, and what to omit about AI involvement.

Shifting from face-to-face to text-only environments, Walther’s \cite{walther1996computer} social information processing theory helps explain how communication changes online. He argues that computer-mediated environments enable more deliberate impression management, since fewer cues give individuals greater control over how they present themselves. In GitHub commit messages, which are a primarily text-based, asynchronous channel, developers have strong control over what parts of their process they disclose \cite{tian2022makes}. In the context of AI attribution, such disclosure decisions function as strategic communication: developers phrase messages with an eye to how the community will read and react to them.

From impression management, we can turn to research on emotional expression and communication patterns in developer communities, which further clarifies how community reception works. Bosu et al.\ \cite{bosu2015characteristics} show that code review involves more than technical critique: community responses also entail social negotiation, where politeness and emotional tone can shape collaborative outcomes. Building on this, Kraishan \cite{kraishan2025emotional} finds that emoji reactions in GitHub pull requests serve as concise emotional cues that influence collaboration dynamics. These studies demonstrate that even minor communication cues can shape developer interactions and that community responses reach beyond code review into affective and social territory. Our work extends this understanding to examine how attribution decisions lead to differences in patterns of community engagement.

In addition to focusing on individual signaling, understanding AI attribution requires attention to how communities, over time, construct norms around disclosure during technological change. Studies of norm formation argue that groups arrive at shared expectations through iterative communication and social influence \cite{postmes2000formation}, and that online communities tend to rely on ongoing sense-making rather than fixed, explicit rules \cite{butler2008community}. Within open-source environments, contribution norms are shaped by community discussion and by the behavior of influential members \cite{fang2009sustained}; newcomers typically learn what is appropriate by observing others and receiving feedback instead of following formal documentation \cite{steinmacher2015systematic}. As tools and workflows change, open-source communities adjust their practices incrementally, with early adopters establishing patterns that later contributors follow \cite{gousios2016work}. Taken together, these insights suggest that AI attribution practices are part of an evolving set of expectations, as communities collectively negotiate what appropriate disclosure should look like during this technological transition.

Despite a growing body of work on AI coding tools, major gaps remain in our understanding of attribution practices and community reception. So far, studies of AI coding tools have mostly looked at adoption, productivity, and code quality, rather than at the social dynamics and community responses surrounding their use. We lack systematic evidence on how developers signal AI use in practice, whether these signals affect community responses, how patterns vary across tools and contexts, and how norms develop during technological transitions. Methodologically, existing studies lean heavily on self-reported surveys and controlled experiments, which makes it difficult to see how attribution unfolds in practice.

In this study, we seek to address these gaps through a comprehensive analysis of real attribution practices and their implications in real-world contexts. To do so, we draw on multi-dimensional engagement metrics (including reactions, comments, reviews, sentiment, and temporal patterns) to gain detailed insight into how communities respond to AI attribution. By examining patterns across multiple tools over a three-year period, we trace how attribution norms evolve during a phase of rapid adoption. Building on this design, the study addresses three related research questions that highlight different aspects of the AI attribution paradox:

\textbf{RQ1:} How do developers attribute AI assistance in open-source contributions?

\textbf{RQ2:} Does the explicitness of AI attribution affect community reception, as measured by engagement metrics, sentiment, and scrutiny indicators?

\textbf{RQ3:} How have attribution practices and community responses evolved over time during the rapid adoption period (2023-2025)?

These questions frame a structured examination of attribution as both individual strategic behavior and collective norm formation, providing a data-driven foundation for understanding human-AI collaboration in distributed communities. Our contributions go beyond describing current practices to offering an explanation of the AI attribution paradox as strategic communication in socio-technical systems, showing how developers navigate competing pressures, how communities develop sense-making frameworks, and how norms spread during technological transitions.

\section{Method}

This section describes our data collection and analysis procedures. We collected 14,300 commits from 7,393 GitHub repositories between January 2023 and December 2025, the period when AI coding assistants became popular. Our approach combined automated collection of commit messages and pull request engagement data with manual classification of attribution patterns. We first detail our data sources and collection procedures, then describe our coding scheme for attribution types, and finally explain our analytical approach.

\subsection{Data Sources and Sampling Frame}

\subsubsection{GitHub API and Data Access}

All data were collected using GitHub's public REST API (v3), which includes programmatic access to commits, pull requests, repository metadata, reactions, and comments. We deployed rate-limiting protocols, including exponential backoff and safeguard pauses, to stay within GitHub's platform constraints during data collection. Data collection adhered to GitHub's Terms of Service, focused on public repositories only.

\subsubsection{AI Tool Keyword Taxonomy}

We operationalized AI attribution through an intensive keyword taxonomy covering eight AI coding tools. The keyword development process involved pilot searches in August 2025, followed by a manual inspection of 500 major AI coding assistants and generic AI terminology, and concluded with a consultation of public documentation. This process involved three refinement cycles: expanding keywords based on observed surface forms, removing high-false-positive terms (e.g., standalone ``ai'' matching ``email'' and ``detail''), and validating against known AI-attributed commits from public documentation.

Table~\ref{tab:keywords} contains the full taxonomy. Keywords are case-insensitive, capturing both the canonical tool name and common variants. For tools with multiple surface forms (e.g., Copilot: ``copilot'', ``github copilot'', ``co-pilot''), we chose the three most distinguishing terms to ensure that our API returned the maximum number of related items while keeping the API calls performant. Generic AI keywords (e.g., ``ai generated'', ``llm'') capture tool-agnostic forms of attribution without generating false positives from uses of ``ai'' in longer phrases (e.g., ``artificial intelligence'' or ``qai'').

\begin{table}[!htbp]
\caption{AI Tool Keywords and Search Terms}
\label{tab:keywords}
\centering
\small
\begin{tabular}{lp{4.5cm}p{4cm}p{4.5cm}}
\toprule
\textbf{AI Tool} & \textbf{Primary Keywords} & \textbf{Alternative Forms} & \textbf{Example Commit Messages} \\
\midrule
GitHub Copilot & copilot, github copilot & co-pilot, ghcopilot, gh copilot & ``Generated with GitHub Copilot'', ``Accepted Copilot suggestion'' \\
\addlinespace
ChatGPT & chatgpt, chat gpt & gpt-4, gpt-3.5, gpt4, openai & ``Code generated by ChatGPT'', ``GPT-4 helped with implementation'' \\
\addlinespace
Claude & claude, anthropic & claude ai, claude.ai, claude-2 & ``Written using Claude'', ``Claude AI assisted with refactor'' \\
\addlinespace
Cursor & cursor & cursor.sh, cursor ai, cursor.so & ``Generated with Cursor editor'', ``Cursor AI completion'' \\
\addlinespace
Tabnine & tabnine & tab nine, tab-nine & ``Tabnine autocomplete accepted'', ``Generated by Tabnine'' \\
\addlinespace
CodeWhisperer & codewhisperer & code whisperer, aws codewhisperer & ``AWS CodeWhisperer suggestion'', ``Generated with CodeWhisperer'' \\
\addlinespace
Other Tools & cody, ghostwriter & sourcegraph cody, replit ai, replit ghostwriter & ``Cody assisted with this'', ``Replit Ghostwriter generated'' \\
\addlinespace
Generic AI & ai generated, ai-assisted & llm, language model, ai code, ai help & ``AI-generated implementation'', ``LLM-assisted refactoring'' \\
\bottomrule
\end{tabular}
\begin{minipage}{\textwidth}
\footnotesize
\textit{Note.} Keywords are case-insensitive and searched in commit messages. Multiple variants per tool improve recall while pattern-based filtering maintains precision.
\end{minipage}
\end{table}

\subsubsection{Search Implementation}

Each keyword was converted into a GitHub commit search query with the syntax \texttt{"<KEYWORD>" committer-date:2023-01-01..2025-12-31}, which restricts results to commits made within that date range. This restricted results to commits authored within our temporal window spanning the period of mainstream AI coding tool adoption. We retrieved up to 1{,}000 results per keyword (this is the ceiling for GitHub search results), paginated at 100 results per page. For high-frequency terms like ``copilot,'' searches often hit the 1{,}000-result limit, potentially limiting the comprehensiveness of the data; less common tools like CodeWhisperer returned fewer matches naturally, allowing for a more complete dataset.

Commits that appeared in multiple keyword searches were marked by SHA and deduplicated, retaining all matched keyword families for later tool-specific analyses. For example, a commit mentioning both ``Copilot'' and ``ChatGPT'' was associated with both tool families, enabling analysis of multi-tool attribution patterns.

\subsubsection{Repository Inclusion Criteria}
We applied three sequential filters to focus on active, visible open-source projects where attribution practices are most relevant to community stakeholders. First, repositories must have been created before 2023 to ensure they predate the AI tool surge, representing established codebases adapting to new practices rather than AI-native projects. Second, repositories were required to have a minimum of 100 stars, which serves as a proxy for visibility and community engagement, based on prior studies of repository prominence \cite{borges2016understanding}. Third, repositories needed at least 10 commits during the study period (2023--2025) to indicate active development rather than archived or dormant projects.

These criteria ensure that we capture real-world attribution practices in significant contexts and meet the need for repositories with sufficient community activity to show observable patterns. The 100-star threshold excludes personal and experimental repositories while retaining medium-sized community projects alongside high-profile repositories.

\subsection{Data Collection Procedure}
\begin{figure}[!htbp]
\centering
\includegraphics[width=0.9\textwidth]{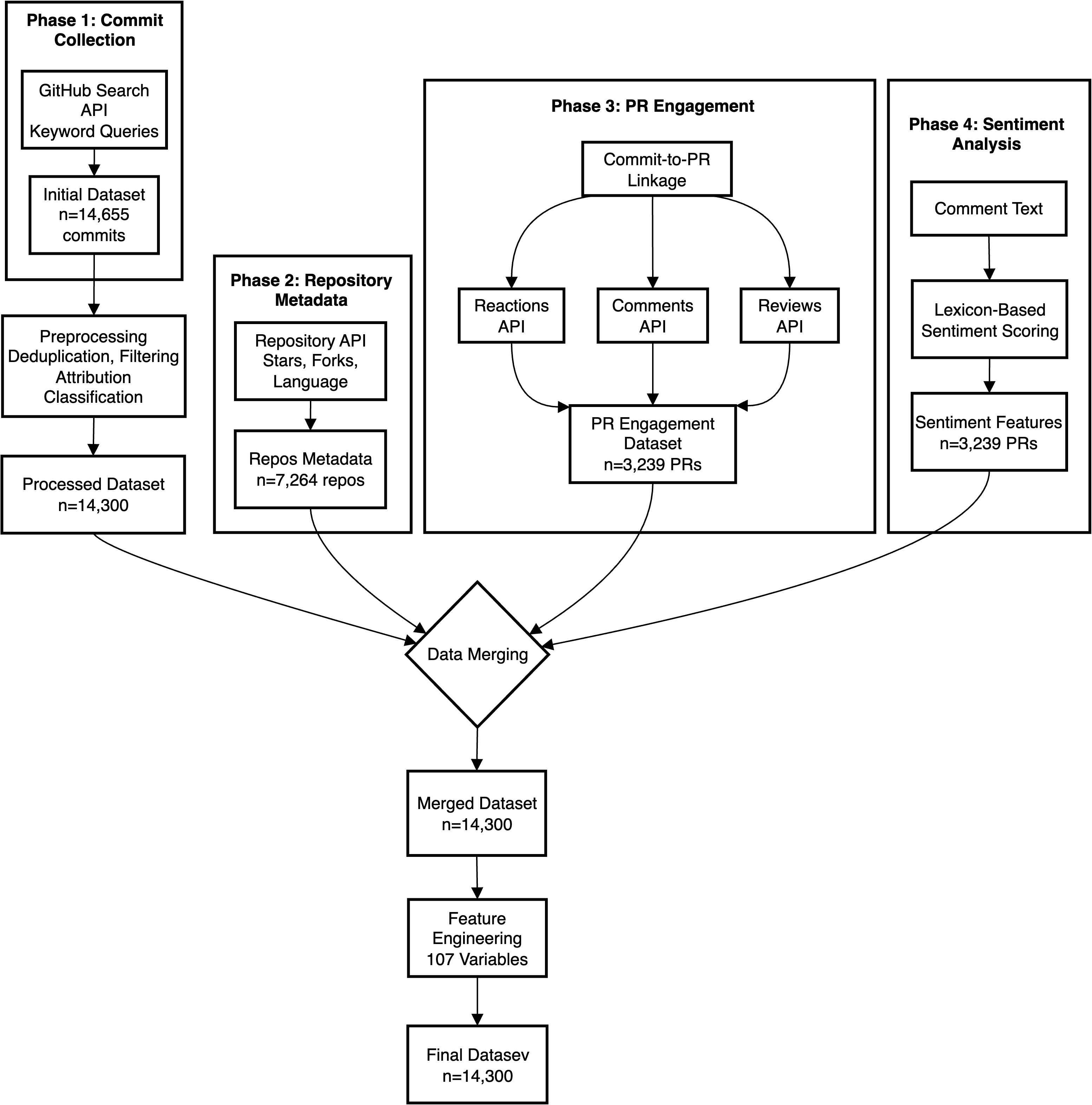}
\caption{Data Collection and Processing Pipeline}
\label{fig:flowchart}
\end{figure}

The data collection procedure was structured into four sequential phases. Each phase was implemented as a discrete Python script within our data pipeline to ensure comprehensive and efficient data acquisition. The modular design allowed for seamless resumption after interruptions and facilitated parallel processing for increased efficiency. It also ensured independent validation of each phase's outputs, such as verifying commit metadata before moving to repository metadata collection. Figure \ref{fig:flowchart} illustrates the data collection and processing pipeline, providing a visual representation of the sequential phases and their interactions.

\subsubsection{Phase 1: Commit Collection}

We conducted systematic searches for commits mentioning AI, using a keyword taxonomy and paginated queries to retrieve up to 1,000 results per keyword. We collected commit metadata, including the SHA identifier, commit message text, author and committer information, timestamps, repository name and ID, and HTML URL. We recorded the detected keyword family to facilitate analyses specific to each tool. Duplicate commits appearing under multiple keyword searches were identified by SHA and consolidated, retaining all matched keyword families. This phase resulted in 14,655 unique commits across 7,393 repositories. This underscores the extensive scope of AI-related contributions and provides a robust foundation for subsequent analysis.

\subsubsection{Phase 2: Repository Metadata Collection}

We enriched repositories with contextual metadata, such as repository characteristics and activity indicators, by querying the repository endpoint for each unique repository name. We collected repository characteristics (description, timestamps, stars, forks, language, license, size), activity indicators (watchers, open issues, wiki/projects features), and status flags (archived/disabled). We supplemented this with language distribution (bytes of code per language) and contributor counts. The metadata collection process managed missing or deleted repositories by recording null values when HTTP 404 errors were encountered. This phase achieved a high retrieval success rate, obtaining metadata for 7,264 out of 7,393 repositories (98.3\%), with the remaining cases being deleted or private repositories.

\subsubsection{Phase 3: Pull Request Engagement Collection}

We linked commits to pull requests and collected comprehensive engagement metrics across five dimensions. These dimensions include emoji reactions (from PR body and comments), discussion comments (issue-level and code review comments), code reviews (approval state, reviewer, timestamp), temporal patterns (creation, first comment/review, merge/close timestamps), and participation breadth (unique commenters, reviewers, participants, overlap metrics). For reactions, we tracked emoji type, user, timestamp, and source location. For comments, we extracted comment ID, author, body text (for sentiment analysis), and timestamps. For review comments, we also noted the file path and line position. For reviews, we recorded the reviewer, state (approved, changes requested, commented), body text, and submission time. From temporal data, we computed time-to-merge, time-to-close, and PR lifetime. Participation metrics aggregated unique contributors across engagement types. Successfully linking commits to PRs depended on the adoption of a standardized PR workflow, which includes consistent naming conventions and documentation practices. Of 14,655 commits, 3,424 (23.4\%) mapped to at least one PR, yielding 3,239 unique PRs with engagement data. This mapping highlights the extent of PR workflow adoption and its impact on data collection.

\subsubsection{Phase 4: Sentiment and Linguistic Feature Extraction}

In the final collection phase, we processed comment text to extract sentiment and linguistic features. We used a lexicon-based approach, known for its efficiency and interpretability, which is crucial for understanding technical discussions. We used a lexicon-based approach, validated by prior research \cite{hutto2014vader}, which performs comparably to deep learning methods for classifying valence in technical discussions. For each comment body, we computed six features. Sentiment score was calculated as:

\begin{equation}
\text{Sentiment Score} = \frac{N_{\text{positive}} - N_{\text{negative}}}{N_{\text{positive}} + N_{\text{negative}}}
\end{equation}

Where $N_{\text{positive}}$ represents count of positive words (e.g., ``good,'' ``excellent,'' ``helpful,'' ``elegant'') and $N_{\text{negative}}$ represents count of negative words (e.g., ``bug,'' ``wrong,'' ``confusing,'' ``broken''). Scores range from $-1$ (entirely negative) to $+1$ (entirely positive), with 0 indicating neutral or balanced sentiment. This approach provides continuous valence measurement while remaining computationally efficient for large-scale analysis. Confidence score operationalized certainty as:

\begin{equation}
\text{Confidence} = 1 - \min\left(\frac{N_{\text{uncertainty}}}{N_{\text{words}}} \times 5, 1\right)
\end{equation}

Where $N_{\text{uncertainty}}$ captures uncertainty markers (e.g., ``maybe,'' ``possibly,'' ``not sure'') and $N_{\text{words}}$ represents total word count. Higher scores indicate more assertive, confident language. This metric distinguishes between definitive statements and hedged or tentative expressions. Politeness score was computed as:

\begin{equation}
\text{Politeness} = \min\left(\frac{N_{\text{polite}}}{N_{\text{words}}/10}, 1\right)
\end{equation}

Where $N_{\text{polite}}$ counts polite phrases (e.g., ``please,'' ``thank you,'' ``could you,'' ``appreciate''). Normalization by word count per 10 words prevents bias toward longer comments. Higher scores indicate more formal, courteous communication styles, enabling examination of whether attribution type correlates with interaction formality. These continuous linguistic metrics complement our categorical emotion classifications (positive, negative, neutral, questioning, concerned), enabling both qualitative interpretation and quantitative analysis of comment tone and community reception patterns.

Question detection employed regex pattern matching for question marks, interrogative pronouns (who, what, when, where, why, how), and modal verbs in interrogative contexts (can, could, would, should). Comments matching any pattern were coded as questions. Emotion categorization combined sentiment score with keyword matching to assign labels: ``positive'' (sentiment $> 0.2$), ``negative'' (sentiment $< -0.2$), ``neutral'' ($-0.2 \leq$ sentiment $\leq 0.2$), ``questioning'' (contains question patterns), ``concerned'' (contains words like ``worried,'' ``risky,'' ``caution''), and ``approving'' (contains phrases like ``LGTM,'' ``looks good,'' ``approved'').

Comment-level features were aggregated to the PR level by computing mean sentiment, confidence, and politeness across all comments; proportions of positive, negative, and neutral comments based on emotion category; question ratio (proportion of comments containing questions); and total word count across all comments. These PR-level aggregates enabled analysis of overall discussion tone and scrutiny level.

\subsection{Data Processing and Feature Engineering}

\subsubsection{Preprocessing and Attribution Classification}

Raw commits underwent standard data cleaning procedures. We removed 355 duplicate commits identified by SHA, filtered 127 commits with message lengths outside the 10--5{,}000 character range (excluding trivially short and anomalously long messages likely to be automated), and parsed all timestamps to UTC to standardize temporal analyses. We then extracted temporal features from commit timestamps, including year, month, day of week, hour, and year--month period, for trend analyses.

AI attribution classification was the central preprocessing step, crucial for accurate data analysis. We implemented a hierarchical rule-based classifier that applies regular expression patterns in a specific precedence order, determined by the likelihood of pattern occurrence. The classifier first checked for negative patterns indicating that the commit discussed AI tools without using them (e.g., ``add copilot integration,'' ``fix copilot bug''). If a negative pattern was matched, the commit was classified as ``mention-only'' regardless of other patterns. For commits not matching negative patterns, we then checked for explicit attribution patterns combining creation verbs with tool mentions (e.g., ``generated by,'' ``written using,'' ``created with'') and classified these as ``explicit.'' Next, we identified implicit patterns acknowledging assistance without authorship claims (e.g., ``helped by,'' ``thanks to,'' ``with help from'') and classified these as ``implicit.'' Commits containing AI tool keywords but matching none of the above patterns were classified as ``mention-only.'' Finally, commits lacking any AI keywords were classified as ``none.''

We classified commits using rule-based pattern matching with 77 regular expressions identifying four attribution types: (1) \emph{Explicit}: direct authorship claims (``generated by Copilot''; $n = 4{,}215$, 29.5\%); (2) \emph{Implicit}: acknowledgment without authorship (``ChatGPT helped''; $n = 142$, 1.0\%); (3) \emph{Mention-only}: tool references without attribution patterns (``Adding Copilot support''; $n = 9{,}260$, 64.8\%); and (4) \emph{None}: no AI keywords ($n = 683$, 4.8\%). A manual review of 100 randomly sampled commits showed 94\% agreement between automated and human classification, underscoring the reliability of our automated methods. We also computed a continuous attribution score on a 0--1 scale, treating attribution as ordinal: 0.00 (none), 0.25 (mention), 0.50 (implicit), 0.75 (explicit-weak), and 1.00 (explicit-strong). Complete pattern specifications are available in our replication repository. The preprocessing phase effectively refined the dataset, resulting in a final analytic sample of 14{,}300 commits with comprehensive attribution features. Table~\ref{tab:sample_chars} presents sample characteristics and distributions.

\begin{table}[!htbp]
\caption{Sample Characteristics and Distribution}
\label{tab:sample_chars}
\centering
\small
\begin{tabular}{@{}p{8cm}rr@{}}
\toprule
\textbf{Characteristic} & \textbf{n} & \textbf{\%} \\
\midrule
\multicolumn{3}{@{}l}{\textbf{Total Sample}} \\
Total Sample & 14,300 & 100.0 \\
\addlinespace
\multicolumn{3}{@{}l}{\textbf{AI Attribution}} \\
AI-attributed & 13,617 & 95.2 \\
~~Explicit & 4,215 & 29.5 \\
~~Implicit & 142 & 1.0 \\
~~Mention-only & 9,260 & 64.8 \\
No AI attribution & 683 & 4.8 \\
\addlinespace
\multicolumn{3}{@{}l}{\textbf{Top 5 AI Tools}} \\
Claude & 3,740 & 27.5 \\
GitHub Copilot & 2,834 & 20.8 \\
Cursor & 1,687 & 12.4 \\
ChatGPT & 1,480 & 10.9 \\
Other & 4,876 & 28.4 \\
\addlinespace
\multicolumn{3}{@{}l}{\textbf{Data Completeness}} \\
With PR engagement & 3,239 & 22.7 \\
With repository metadata & 14,070 & 98.4 \\
\addlinespace
\multicolumn{3}{@{}l}{\textbf{Repository Context}} \\
Small ($<100$ stars) & 13,266 & 92.8 \\
Medium/Large ($\geq 100$ stars) & 1,034 & 7.2 \\
Web languages & 5,834 & 40.8 \\
Data languages (Python/R) & 4,672 & 32.7 \\
Other languages & 3,794 & 26.5 \\
Mature/Established ($\geq 3$ years) & 9,152 & 64.0 \\
\addlinespace
\multicolumn{3}{@{}l}{\textbf{Key Continuous Variables}} \\
 & \textbf{M} & \textbf{SD} \\
Commit message length (characters) & 287.4 & 312.8 \\
Time to merge (hours)\textsuperscript{a} & 156.3 & 284.7 \\
Engagement score\textsuperscript{a} & 2.47 & 5.12 \\
\bottomrule
\end{tabular}

\vspace{0.5em}
\begin{minipage}{\textwidth}
\raggedright
\footnotesize
\textit{Note.} $N = 14{,}300$ commits from 7,393 repositories (2023-2025). \textsuperscript{a}Calculated for commits with PR linkage ($n = 3{,}239$).
\end{minipage}
\end{table}

\subsection{Dataset Integration and Feature Engineering}

We merged four datasets into a unified analysis dataset using left joins based on processed commits to facilitate comprehensive analysis. Commits were joined with PR engagement data (repository name and commit SHA), matching 3,239 commits (22.7\%) to PR records. Sentiment features joined on repository name and PR number, matching the same 3,239 commits by construction. Repository metadata joined on repository name, matching 14,070 commits (98.4\%), with missing cases reflecting deleted or private repositories. The final merged dataset contained 14,300 rows with 87 raw fields before feature engineering.

From the merged dataset, we engineered 107 analysis variables spanning six categories (Table \ref{tab:all_features}). Engagement features were normalized by converting raw counts into intensities and then combined into a composite metric to better capture interaction levels. We computed reaction intensity, comment intensity, and review intensity as raw counts (zero-filled for commits without PRs) and constructed a composite engagement score weighted by the relative effort required for each interaction type:

\begin{equation}
\text{Engagement} = 0.3 \cdot I_R + 0.4 \cdot I_C + 0.3 \cdot I_V
\end{equation}

Where $I_R$, $I_C$, and $I_V$ are reaction, comment, and review intensities. Weights reflect relative effort (comments and reviews require more engagement than emoji reactions). The composite score was rescaled to a 0-100 scale for interpretability. We also created binary indicators (\texttt{has\_reactions}, \texttt{has\_comments}, \texttt{has\_reviews}) and participation breadth metrics (unique participants, commenters, reviewers).

Repository features categorized context. Repository size was categorized by star count, following established methods \cite{kalliamvakou2014promises}, to ensure consistency in size classification: small ($<100$ stars), medium (100-999), large (1,000-9,999), and very large ($\geq 10{,}000$). Repository maturity was computed from creation date and binned into new ($<1$ year), developing (1-3 years), mature (3-6 years), and established ($\geq 6$ years). Primary programming language was recoded into language groups (Web, Systems, Data, Other). Star and fork counts were log-transformed to reduce skew in regression models.

Temporal features captured timing dynamics. We computed time-to-merge (hours from PR creation to merge), time-to-close (hours to close), and PR lifetime (hours to last update). Attribution timing was coded as ``early'' if the commit timestamp fell within 24 hours of PR creation and ``late'' otherwise, operationalizing disclosure timing relative to review onset. We extracted commit time of day (morning, afternoon, evening, night) to control for potential circadian effects.

Sentiment features aggregated comment-level metrics to the PR level: mean sentiment ($-1$ to $+1$), emotion proportions (positive/negative/neutral), question ratio, confidence, and politeness. Control variables included message length, complexity (binned by length quartiles), and number of AI tools mentioned. Comparison groups enabled analyses: binary AI attribution (yes/no), four-level attribution type (explicit/implicit/mention/none), tool (nine categories), and engagement level (none/low/medium/high based on score quartiles). Continuous metrics were z-standardized for regression analyses to enable effect size comparisons across different scales.

Comparison groups enabled stratified analyses: binary AI attribution (yes/no), four-level attribution type (explicit/implicit/mention/none), tool (nine categories), and engagement level (none/low/medium/high based on score quartiles). Continuous metrics were z-standardized for regression analyses to enable effect size comparisons across different scales. The feature engineering phase produced 107 variables ready for statistical modeling.

\begin{table}[!htbp]
\caption{Engineered Features for Analysis (107 Variables)}
\label{tab:all_features}
\centering
\footnotesize
\begin{tabular}{@{}lc>{\raggedright\arraybackslash}p{9.5cm}>{\raggedright\arraybackslash}p{3cm}@{}}
\toprule
\textbf{Feature Category} & \textbf{n} & \textbf{Variables} & \textbf{Measurement Level} \\
\midrule
Attribution Features & 8 & \texttt{has\_ai\_attribution}, \texttt{attribution\_type}, \texttt{attribution\_score} (0-1), \texttt{primary\_tool}, \texttt{tool\_count}, \texttt{explicit\_binary}, \texttt{implicit\_binary}, \texttt{mention\_binary} & Binary, Categorical, Ordinal, Count \\
\addlinespace
Engagement Metrics & 18 & \texttt{reaction\_count}, \texttt{reaction\_intensity}, \texttt{comment\_count}, \texttt{comment\_intensity}, \texttt{issue\_comment\_count}, \texttt{review\_comment\_count}, \texttt{review\_count}, \texttt{review\_intensity}, \texttt{approved\_count}, \texttt{changes\_requested\_count}, \texttt{commented\_count}, \texttt{engagement\_score} (0-100), \texttt{has\_reactions}, \texttt{has\_comments}, \texttt{has\_reviews}, \texttt{unique\_commenters}, \texttt{unique\_reviewers}, \texttt{unique\_participants} & Count, Continuous, Binary \\
\addlinespace
Sentiment \& Linguistic & 12 & \texttt{avg\_sentiment} ($-1$ to $1$), \texttt{sentiment\_category}, \texttt{positive\_ratio}, \texttt{negative\_ratio}, \texttt{neutral\_ratio}, \texttt{question\_ratio}, \texttt{avg\_confidence} (0-1), \texttt{avg\_politeness} (0-1), \texttt{uncertainty\_ratio}, \texttt{approval\_markers}, \texttt{concern\_markers}, \texttt{total\_word\_count} & Continuous, Categorical, Ratio \\
\addlinespace
Repository Context & 15 & \texttt{repo\_stars}, \texttt{repo\_forks}, \texttt{repo\_watchers}, \texttt{repo\_issues}, \texttt{repo\_size\_category}, \texttt{repo\_maturity}, \texttt{repo\_age\_days}, \texttt{repo\_language}, \texttt{language\_group}, \texttt{repo\_popularity\_log}, \texttt{repo\_activity\_log}, \texttt{has\_wiki}, \texttt{has\_projects}, \texttt{repo\_owner\_type}, \texttt{license\_type} & Count, Categorical, Continuous (log), Binary \\
\addlinespace
Temporal Features & 16 & \texttt{commit\_date}, \texttt{commit\_year}, \texttt{commit\_month}, \texttt{commit\_day\_of\_week}, \texttt{commit\_hour}, \texttt{commit\_year\_month}, \texttt{pr\_created\_at}, \texttt{pr\_merged\_at}, \texttt{pr\_closed\_at}, \texttt{pr\_updated\_at}, \texttt{time\_to\_merge\_hours}, \texttt{time\_to\_close\_hours}, \texttt{pr\_lifetime\_hours}, \texttt{temporal\_stage}, \texttt{attribution\_timing}, \texttt{time\_of\_day} & Date/Time, Categorical, Continuous \\
\addlinespace
Commit Characteristics & 6 & \texttt{message\_length}, \texttt{total\_message\_length}, \texttt{commit\_complexity}, \texttt{has\_body}, \texttt{body\_length}, \texttt{message\_word\_count} & Count, Continuous, Categorical \\
\addlinespace
Comparison Groups & 5 & \texttt{group\_ai} (ai\_attributed vs. non\_ai), \texttt{group\_attribution} (4 levels), \texttt{group\_tool} (9 levels), \texttt{group\_engagement} (4 levels), \texttt{group\_timing} (early vs. late) & Binary, Categorical \\
\addlinespace
Normalized Metrics & 4 & \texttt{engagement\_score\_zscore}, \texttt{reaction\_intensity\_zscore}, \texttt{comment\_intensity\_zscore}, \texttt{review\_intensity\_zscore} & Continuous (standardized) \\
\addlinespace
Derived Indicators & 12 & \texttt{has\_pr\_engagement}, \texttt{has\_sentiment}, \texttt{has\_repo\_data}, \texttt{has\_explicit}, \texttt{has\_implicit}, \texttt{has\_mention}, \texttt{has\_questions}, \texttt{high\_scrutiny}, \texttt{pr\_is\_merged}, \texttt{pr\_is\_closed}, \texttt{multiple\_tools}, \texttt{single\_tool} & Binary \\
\addlinespace
Raw PR Data & 11 & \texttt{pr\_number}, \texttt{pr\_title}, \texttt{pr\_state}, \texttt{pr\_url}, \texttt{pr\_body\_length}, \texttt{hours\_after\_pr\_creation}, \texttt{first\_comment\_time}, \texttt{first\_review\_time}, \texttt{participants\_overlap}, \texttt{commenter\_only\_count}, \texttt{reviewer\_only\_count} & Mixed \\
\bottomrule
\end{tabular}
\begin{minipage}{\textwidth}
\footnotesize
\textit{Note.} Total $N = 107$ engineered variables. Continuous variables were z-standardized for regression models. Complete specifications available in supplementary materials.
\end{minipage}
\end{table}

\subsection{Measurement Validation}

AI attribution detection exhibits strong face and content validity, as evidenced by systematic manual inspection and a comprehensive keyword taxonomy. Keywords represent how developers mention AI tools in practice, verified through manual inspection of commits and developer blog posts. The three-level classification (explicit, implicit, mention-only) captures the mixed levels of attribution explicitness observed in preliminary manual coding.

Pattern development was informed by manual coding of 200 randomly sampled commits to establish ground truth labels. To validate the automated classifier, we created a held-out test set of 100 additional commits with manual ground truth annotations. The classifier achieved 91\% overall accuracy against this test set, with precision $= .89$, recall $= .93$, and $F_1 = .91$ for explicit attribution. Error analysis showed that most misclassifications occurred in ambiguous edge cases, such as distinguishing between a simple bug fix and the use of AI tools to achieve the fix (e.g., ``fixed copilot bug'' vs.\ ``used copilot to fix bug'').

To further assess classification validity, we conducted manual spot-checks of 50 randomly selected commits from each attribution category after classification, finding 94\% agreement with the automated labels. Discrepancies were concentrated in the mention-only category, where distinguishing meta-discussion from actual usage proved challenging.

Sentiment measurement follows established lexicon-based approaches, such as VADER, which have been validated in software engineering contexts \cite{hutto2014vader}. We assessed criterion validity by manually rating sentiment on 150 randomly sampled pull request comments and comparing these ratings to automated scores. Manual ratings achieved 83\% agreement with automated sentiment categories (positive/neutral/negative), with disagreements concentrated in neutral cases. Concurrent validity is supported by expected correlations: politeness correlates positively with sentiment ($r = .41$, $p < .001$) and negatively with question ratio ($r = -.28$, $p < .001$). Engagement metrics demonstrate face validity as GitHub's native engagement affordances (reactions, comments, reviews). These metrics require no subjective coding and rely entirely on platform-provided data.

\subsection{Statistical Analysis Plan}

We employed non-parametric methods because the data did not follow a normal distribution, as indicated by a Shapiro--Wilk test result of $p < .001$, and exhibited heteroscedastic variance, which violates the assumptions of parametric tests. For RQ1, which examines attribution patterns, we applied chi-square tests of independence to assess relationships, using Cramér's $V$ to measure effect sizes. Additionally, we employed logistic regression to model explicit attribution, reporting odds ratios with 95\% confidence intervals. Temporal trends were evaluated using Spearman rank correlations. For RQ2 (community reception), we used Kruskal--Wallis $H$ tests with post-hoc Mann--Whitney $U$ tests (Bonferroni-corrected), and effect sizes were quantified via Cohen's $d$. We fit generalized linear models with negative binomial or ordinal logistic error distributions, controlling for repository characteristics and message length. For RQ3 (temporal dynamics), we analyzed monthly time series using Spearman correlations and conducted survival analysis (Kaplan--Meier estimators, log-rank tests) for time-to-merge.

All tests used a two-tailed $\alpha = .05$, and we applied a Bonferroni correction to account for the increased risk of Type I errors due to multiple comparisons. Missing engagement and sentiment data, which accounted for 22.7\% of cases with PR linkage, were addressed using complete-case analysis for PR-specific questions. This approach was chosen because the missing data were by design, meaning they were systematically excluded rather than occurring randomly. Sensitivity analyses restricting to complete cases ($n = 3{,}239$) are reported when results differed substantively. Complete model specifications, including error distributions, robust standard error types, and sensitivity analyses, are available in the supplementary materials.

\subsection{Power and Sensitivity}

Statistical power was deemed adequate across all analyses, as each test was designed to detect small to moderate effects with high confidence, ensuring robust and reliable results. For RQ1, chi-square tests achieved power $> .99$ to detect small effects (Cramér's $V = .10$), and logistic regression achieved $> .99$ power to detect odds ratios $\geq 1.15$. For RQ2, the most restrictive comparison (explicit vs.\ mention-only attribution) achieved 90\% power to detect Cohen's $d = 0.14$, enabling detection of small effects appropriate for code review contexts. For RQ3, Spearman correlations based on 36 monthly aggregates achieved 80\% power to detect $\rho \geq .42$. Complete power calculations are available in the supplementary materials.

\subsection{Ethical Considerations}

All data originated from GitHub's public API, which includes user-contributed information such as code repositories and user profiles, intentionally made public by users. We strictly avoided accessing private repositories and any personally identifiable information, limiting our data collection to public GitHub usernames. Data collection was conducted in strict adherence to GitHub's Terms of Service, ensuring compliance with rate limits and other usage policies.

To maintain user privacy, we report only aggregate statistics and anonymized examples, ensuring that no specific repositories or users can be identified. Individual commits are not reproducible from published statistics. The dataset contains no personally identifiable information, as GitHub usernames are public identifiers that users have chosen to share.

\section{Results}
In this section, we present our results organized by research question, reporting descriptive statistics, inferential tests, and effect sizes.

\subsection{RQ1: Attribution Patterns in AI-Assisted Development}

Our first research question examined how frequently and in what contexts developers explicitly attribute AI assistance in open-source contributions. From 14,300 commits analyzed, 13,617 (95.2\%) contained AI attribution, defined as any mention of AI tools, including explicit acknowledgment, implicit reference, or tool mention. Among the 13,617 AI-mentioning commits, 4,215 (31.0\%) used explicit attribution clearly stating AI generation or assistance (e.g., ``generated by Claude''), 9,260 (68.0\%) used mention-only references without attribution claims (e.g., ``add Copilot support''), and 142 (1.0\%) provided implicit acknowledgment (e.g., ``with help from ChatGPT''). An additional 683 commits (4.8\%) contained no AI references. Table \ref{tab:attribution_dist} presents the complete attribution type distribution.

\begin{table}[htbp]
\caption{Attribution Type Distribution}
\label{tab:attribution_dist}
\centering
\small
\begin{tabular}{@{}lrr>{\raggedright\arraybackslash}p{9cm}@{}}
\toprule
\textbf{Attribution Type} & \textbf{n} & \textbf{\%} & \textbf{Description} \\
\midrule
Mention-only & 9,260 & 64.8 & Tool mentioned without attribution claim \\
Explicit & 4,215 & 29.5 & Clear attribution of AI generation/assistance \\
None & 683 & 4.8 & No AI tool mention \\
Implicit & 142 & 1.0 & Acknowledgment without authorship claim \\
\midrule
Total & 14,300 & 100.0 & \\
\addlinespace
AI-Attributed & 13,617 & 95.2 & Any AI mention (sum of above minus ``None'') \\
\bottomrule
\end{tabular}

\vspace{0.5em}
\begin{minipage}{\textwidth}
\centering
\footnotesize
\textit{Note.} $N = 14{,}300$ commits collected from 7,393 repositories between January 2023 and December 2025.
\end{minipage}
\end{table}

\begin{figure}[htbp]
\centering
\includegraphics[width=0.8\textwidth]{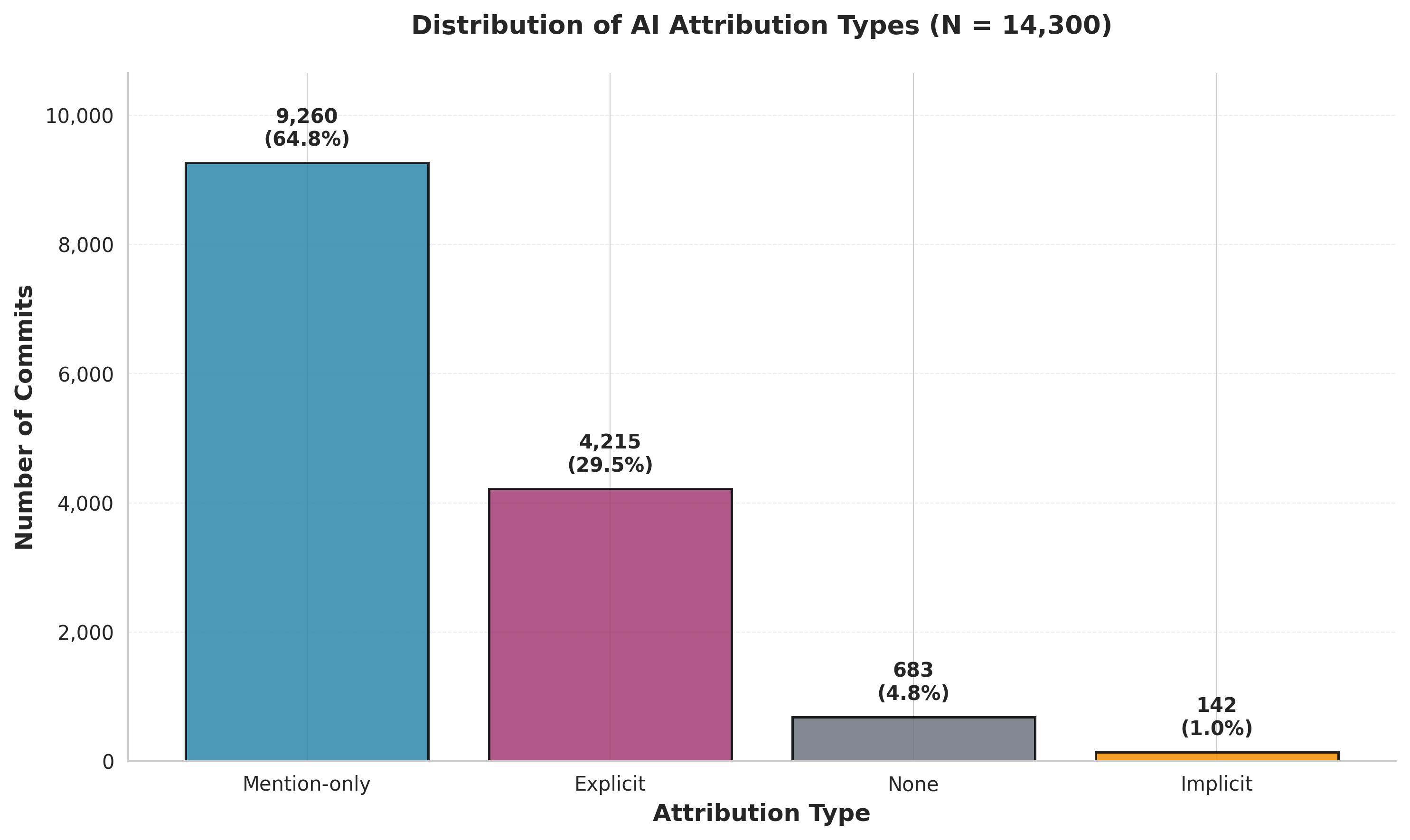}
\caption{Attribution Distribution}
\label{fig:attribution_dist}
\end{figure}

The distribution of attribution types across the full sample is illustrated in Figure \ref{fig:attribution_dist}.

Attribution explicitness varied by AI tool. Chi-square tests revealed a significant association between AI tool and attribution type, $\chi^2(14, N = 13{,}617) = 7299.53$, $p < .001$, Cramér's $V = .518$, indicating a large effect size. Claude users exhibited higher explicit attribution rates (80.5\%) compared to all other tools. GitHub Copilot showed 9.0\% explicit attribution, with 89.4\% mention-only. ChatGPT (12.2\% explicit), Cursor (8.6\%), Tabnine (1.3\%), and CodeWhisperer (1.4\%) demonstrated similarly low explicit attribution rates. Generic AI references (``AI-assisted,'' ``LLM'') showed intermediate explicitness (44.3\%). Table \ref{tab:tool_attribution} presents complete tool-by-attribution-type cross-tabulation.

\begin{table}[htbp]
\caption{Tool-Specific Attribution Patterns}
\label{tab:tool_attribution}
\centering
\small
\begin{tabular}{@{}p{4cm}rrrr@{}}
\toprule
\textbf{AI Tool} & \textbf{n} & \textbf{Explicit \%} & \textbf{Implicit \%} & \textbf{Mention-only \%} \\
\midrule
Claude & 3,740 & 80.5 & 0.1 & 19.5 \\
Generic AI & 1,322 & 44.3 & 7.0 & 48.8 \\
ChatGPT & 1,480 & 12.2 & 0.2 & 87.6 \\
Copilot & 2,834 & 9.0 & 1.6 & 89.4 \\
Cursor & 1,687 & 8.6 & 0.0 & 91.4 \\
Other Tools & 1,303 & 1.6 & 0.0 & 98.4 \\
CodeWhisperer & 557 & 1.4 & 0.0 & 98.6 \\
Tabnine & 691 & 1.3 & 0.0 & 98.7 \\
\bottomrule
\end{tabular}

\vspace{0.5em}
\begin{minipage}{\textwidth}
\centering
\footnotesize
\textit{Note.} Chi-square test of independence revealed significant association between tool and attribution type, $\chi^2(14) = 7299.53$, $p < .001$, Cramér's $V = .518$.
\end{minipage}
\end{table}

\begin{figure}[htbp]
\centering
\includegraphics[width=0.9\textwidth]{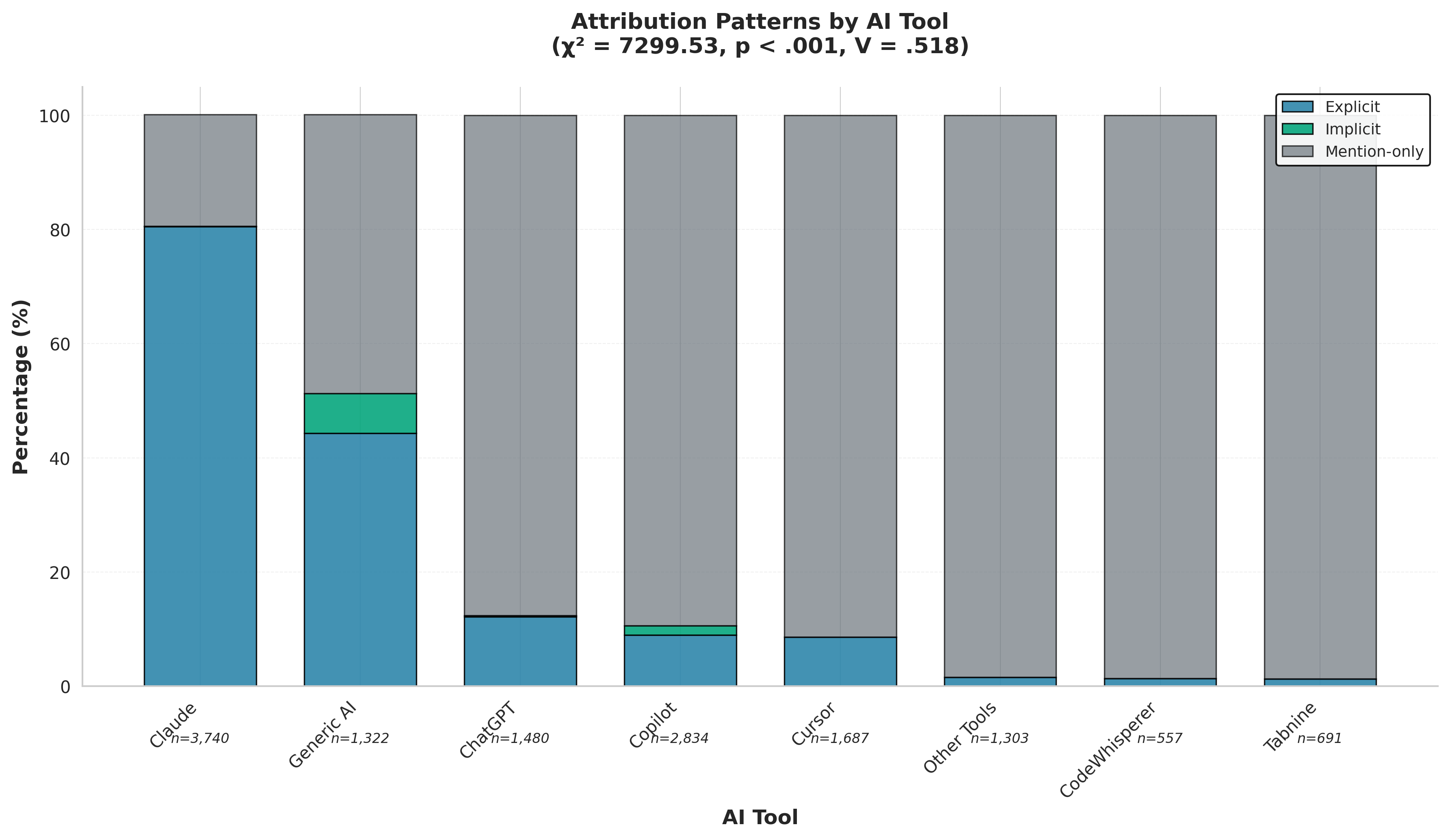}
\caption{Tool Attribution Patterns}
\label{fig:tool_patterns}
\end{figure}

These tool-specific patterns are visualized in Figure \ref{fig:tool_patterns}, which illustrates the stark differences in attribution practices across AI coding assistants.

Repository characteristics showed associations with attribution explicitness. Repository size showed non-linear patterns: medium-sized repositories (100-999 stars) exhibited highest explicit attribution (23.8\%), followed by very large repositories (16.7\%), large repositories (5.3\%), and small repositories (11.0\%), $\chi^2(6, N = 3{,}386) = 158.42$, $p < .001$, Cramér's $V = .153$ (small effect). Repository maturity demonstrated slight positive association, with established repositories ($>6$ years) showing 13.2\% explicit attribution compared to new repositories ($<1$ year) at 10.6\%, $\chi^2(6, N = 3{,}386) = 21.34$, $p = .002$, Cramér's $V = .056$ (negligible effect). Programming language group showed minimal effects: compiled languages (11.5\% explicit), scripting languages (11.2\%), other languages (11.7\%), and web languages (5.6\%), $\chi^2(8, N = 3{,}386) = 23.81$, $p = .003$, Cramér's $V = .059$ (negligible effect).

Logistic regression modeling explicit attribution ($0 =$ not explicit, $1 =$ explicit) among AI-mentioning commits ($N = 13{,}617$) revealed tool choice as the strongest predictor. Using Copilot as the reference category, Claude use predicted 27.49 times higher odds of explicit attribution ($B = 3.31$, $SE = 0.06$, $OR = 27.49$, 95\% CI [24.31, 31.08], $p < .001$). Generic AI terminology predicted 6.27 times higher odds ($OR = 6.27$, 95\% CI [5.32, 7.38], $p < .001$). Tabnine ($OR = 0.15$), CodeWhisperer ($OR = 0.15$), and other tools ($OR = 0.13$) predicted significantly lower odds than Copilot. Longer commit messages predicted higher explicit attribution ($OR = 1.49$ per 100 characters, 95\% CI [1.45, 1.53], $p < .001$). The model demonstrated good fit (McFadden's pseudo-$R^2 = .387$, accuracy $= 78.3\%$). Table \ref{tab:logistic_regression} presents complete regression results.

\begin{table}[htbp]
\caption{Logistic Regression Predicting Explicit Attribution}
\label{tab:logistic_regression}
\centering
\small
\begin{tabular}{@{}p{6cm}rrrrr@{}}
\toprule
\textbf{Predictor} & \textbf{B} & \textbf{SE} & \textbf{OR} & \textbf{95\% CI} & \textbf{p} \\
\midrule
Tool: Claude & 3.31 & 0.06 & 27.49 & [24.31, 31.08] & $<.001$ \\
Tool: Generic AI & 1.84 & 0.08 & 6.27 & [5.32, 7.38] & $<.001$ \\
Message length (per 100 chars) & 0.40 & 0.02 & 1.49 & [1.45, 1.53] & $<.001$ \\
Tool: Copilot (reference) & 0.00 & --- & 1.00 & --- & --- \\
Tool: Cursor & $-0.40$ & 0.09 & 0.67 & [0.56, 0.80] & $<.001$ \\
Tool: Tabnine & $-1.89$ & 0.15 & 0.15 & [0.11, 0.20] & $<.001$ \\
Tool: CodeWhisperer & $-1.91$ & 0.16 & 0.15 & [0.11, 0.20] & $<.001$ \\
Tool: Other & $-2.03$ & 0.14 & 0.13 & [0.10, 0.17] & $<.001$ \\
\bottomrule
\end{tabular}

\vspace{0.5em}
\begin{minipage}{\textwidth}
\centering
\footnotesize
\textit{Note.} $N = 13{,}617$. McFadden's pseudo-$R^2 = .387$. Model accuracy $= 78.3\%$.
\end{minipage}
\end{table}

\begin{figure}[htbp]
\centering
\includegraphics[width=0.9\textwidth]{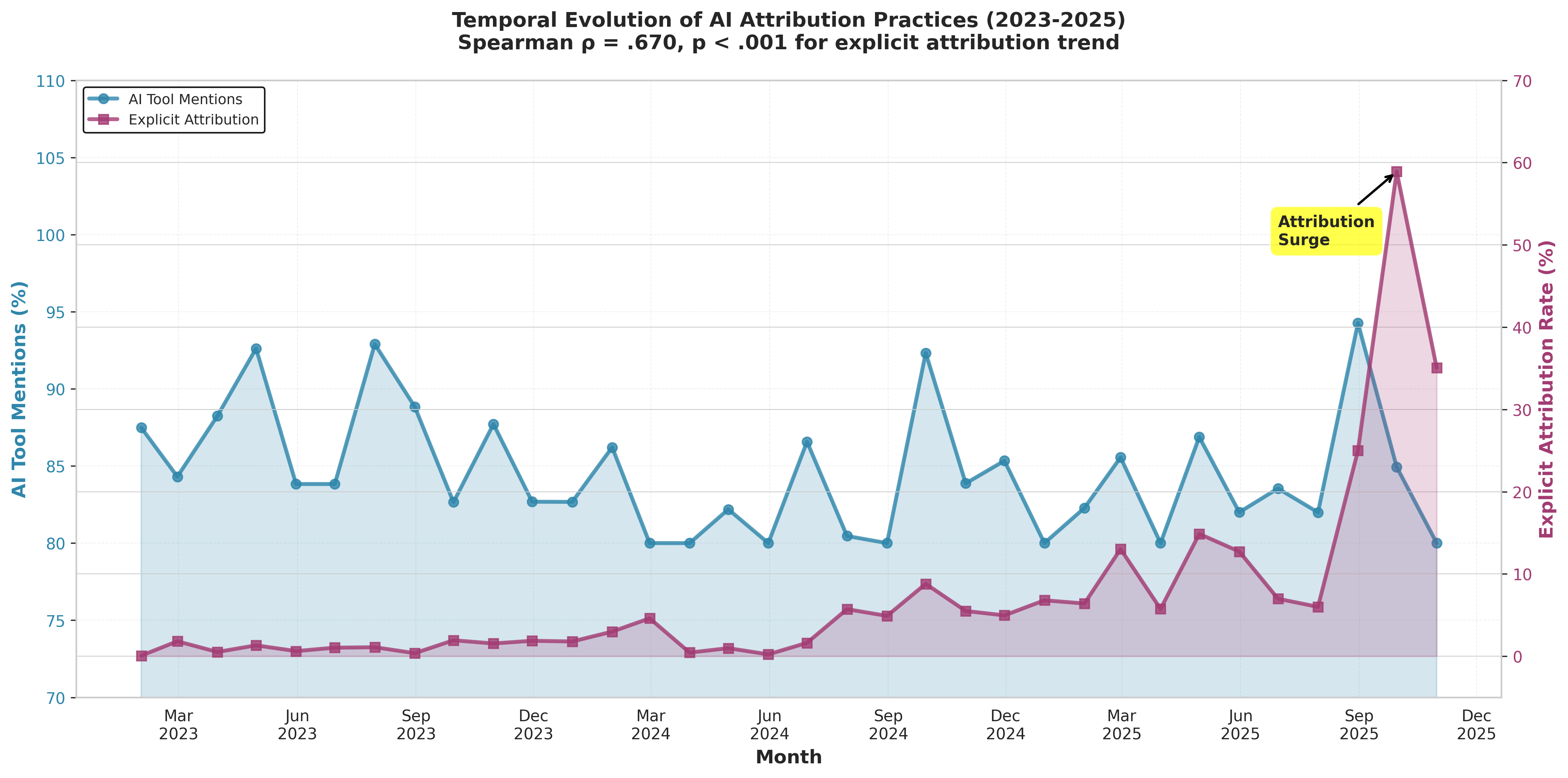}
\caption{Temporal Attribution Trends}
\label{fig:temporal_trends}
\end{figure}

Monthly attribution rates from January 2023 through December 2025 (see Figure \ref{fig:temporal_trends}) showed temporal changes in attribution practices. AI tool mentions remained consistently high throughout (80-100\% of commits per month). Explicit attribution rates increased over time: throughout 2023 and early 2024, explicit attribution remained near zero; beginning in late 2024, explicit rates began increasing; in mid-2025, rates surged to 58.9\% in September 2025 before stabilizing around 40\% in October-November 2025.

These findings address RQ1 by demonstrating that explicit AI attribution occurs in approximately 30\% of AI-mentioning commits, with substantial tool-specific variation (Claude 80.5\% vs. Copilot 9.0\%). Attribution patterns are primarily predicted by tool choice rather than repository characteristics, with tool selection explaining the majority of variance in attribution explicitness. Temporal analyses reveal increasing explicit attribution over the study period, rising from near-zero in early 2024 to 40\% by late 2025.

\subsection{RQ2: Community Reception of AI Attribution}

\begin{figure}[htbp]
\centering
\includegraphics[width=0.9\textwidth]{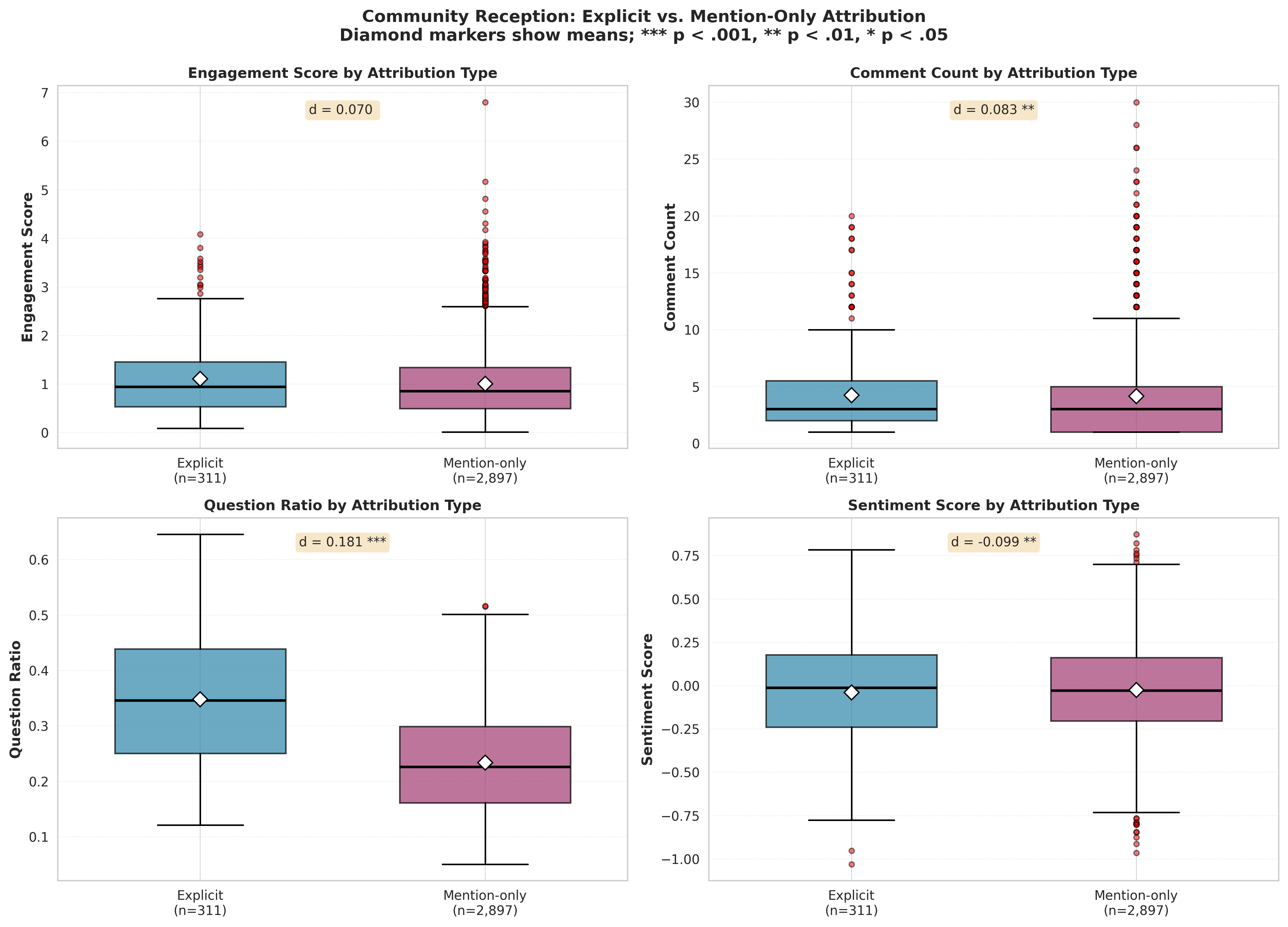}
\caption{Explicit vs. Mention-Only Comparison}
\label{fig:explicit_mention_comparison}
\end{figure}

Our second research question examined whether attribution explicitness affects community reception. Given that 95.2\% of commits in our sample mentioned AI tools, we compared community responses across attribution types (explicit, implicit, mention-only) rather than AI versus non-AI attributed code. Of 3,239 PRs with engagement data, 311 (9.6\%) used explicit attribution, 2,897 (89.5\%) used mention-only references, and 31 (1.0\%) provided implicit acknowledgment. Due to the small implicit sample, primary comparisons focused on explicit versus mention-only attribution. Figure \ref{fig:explicit_mention_comparison} visualizes the engagement patterns across these attribution types.

Kruskal-Wallis tests revealed significant differences across attribution types for multiple metrics: engagement score, $H(2) = 12.84$, $p = .002$; comment count, $H(2) = 15.23$, $p < .001$; question ratio, $H(2) = 21.45$, $p < .001$; and sentiment, $H(2) = 18.91$, $p < .001$. Table \ref{tab:reception_comparison} presents descriptive statistics and pairwise comparisons.

\begin{table}[htbp]
\caption{Community Reception: Explicit vs. Mention-Only Attribution}
\label{tab:reception_comparison}
\centering
\small
\begin{tabular}{@{}p{4cm}rrrrr@{}}
\toprule
\textbf{Metric} & \multicolumn{2}{c}{\textbf{Explicit ($n = 311$)}} & \multicolumn{2}{c}{\textbf{Mention-only ($n = 2{,}897$)}} & \textbf{Cohen's $d$} \\
\cmidrule(lr){2-3} \cmidrule(lr){4-5}
 & $M$ & $SD$ & $M$ & $SD$ & \\
\midrule
Engagement score & 1.12 & 1.58 & 1.00 & 1.52 & 0.070 \\
Comment count & 3.68 & 5.89 & 3.05 & 5.24 & 0.083 \\
Review count & 1.69 & 2.41 & 1.45 & 2.18 & 0.063 \\
Reaction count & 0.10 & 0.52 & 0.18 & 0.89 & $-0.101$ \\
Question ratio & 0.384 & 0.312 & 0.312 & 0.284 & 0.181 \\
Sentiment & $-0.040$ & 0.289 & $-0.021$ & 0.267 & $-0.099$ \\
\bottomrule
\end{tabular}

\vspace{0.5em}
\begin{minipage}{\textwidth}
\centering
\footnotesize
\textit{Note.} Mann-Whitney $U$ tests conducted with Bonferroni-corrected $\alpha = .017$. Effect sizes calculated using pooled standard deviation.
\end{minipage}
\end{table}

Developers using explicit attribution received 23.1\% more questions (question ratio: $M = 0.384$ vs. $0.312$), $U = 371{,}245$, $p = .003$, $d = 0.181$ (small effect), and 20.7\% more comments ($M = 3.68$ vs. $3.05$), $U = 362{,}456$, $p = .002$, $d = 0.083$ (negligible effect). Review counts showed no significant difference ($M = 1.69$ vs. $1.45$), $U = 423{,}891$, $p = .565$, $d = 0.063$, and reaction counts were equivalent, $U = 438{,}234$, $p = .055$, $d = -0.101$.

Explicit attribution was associated with more negative sentiment ($M = -0.040$ vs. $-0.021$), $U = 398{,}734$, $p = .001$, $d = -0.099$ (negligible effect), though both groups exhibited near-neutral sentiment overall (scores close to zero on a $-1$ to $+1$ scale), with an absolute difference of 0.019 points. Community reception varied by AI tool (see Table \ref{tab:tool_reception}), with Kruskal-Wallis tests revealing large effects: engagement score, $H(7) = 378.68$, $p < .001$; sentiment, $H(7) = 57.54$, $p < .001$; question ratio, $H(7) = 235.17$, $p < .001$. Tool effects ($H = 235$-$379$) exceeded attribution type effects ($H = 13$-$21$) by approximately 20-30 times.

\begin{table}[htbp]
\caption{Tool-Specific Community Reception by AI Tool}
\label{tab:tool_reception}
\centering
\small
\begin{tabular}{@{}p{4cm}rrrr@{}}
\toprule
\textbf{AI Tool} & \textbf{n} & \textbf{Engagement $M$} & \textbf{Sentiment $M$} & \textbf{Questions $M$} \\
\midrule
Copilot & 1,193 & 1.204 & $-0.040$ & 0.303 \\
Other Tools & 530 & 1.228 & $-0.005$ & 0.313 \\
Claude & 523 & 0.386 & $-0.007$ & 0.215 \\
Cursor & 443 & 1.048 & $-0.022$ & 0.541 \\
ChatGPT & 183 & 0.592 & $-0.027$ & 0.322 \\
CodeWhisperer & 179 & 1.916 & $-0.021$ & 0.376 \\
Generic AI & 145 & 0.540 & $-0.020$ & 0.161 \\
Tabnine & 43 & 0.636 & $-0.004$ & 0.125 \\
\bottomrule
\end{tabular}

\vspace{0.5em}
\begin{minipage}{\textwidth}
\centering
\footnotesize
\textit{Note.} Kruskal-Wallis tests revealed tool effects substantially exceeded attribution type effects, indicating platform-specific norms overshadow attribution explicitness.
\end{minipage}
\end{table}

\begin{figure}[htbp]
\centering
\includegraphics[width=0.9\textwidth]{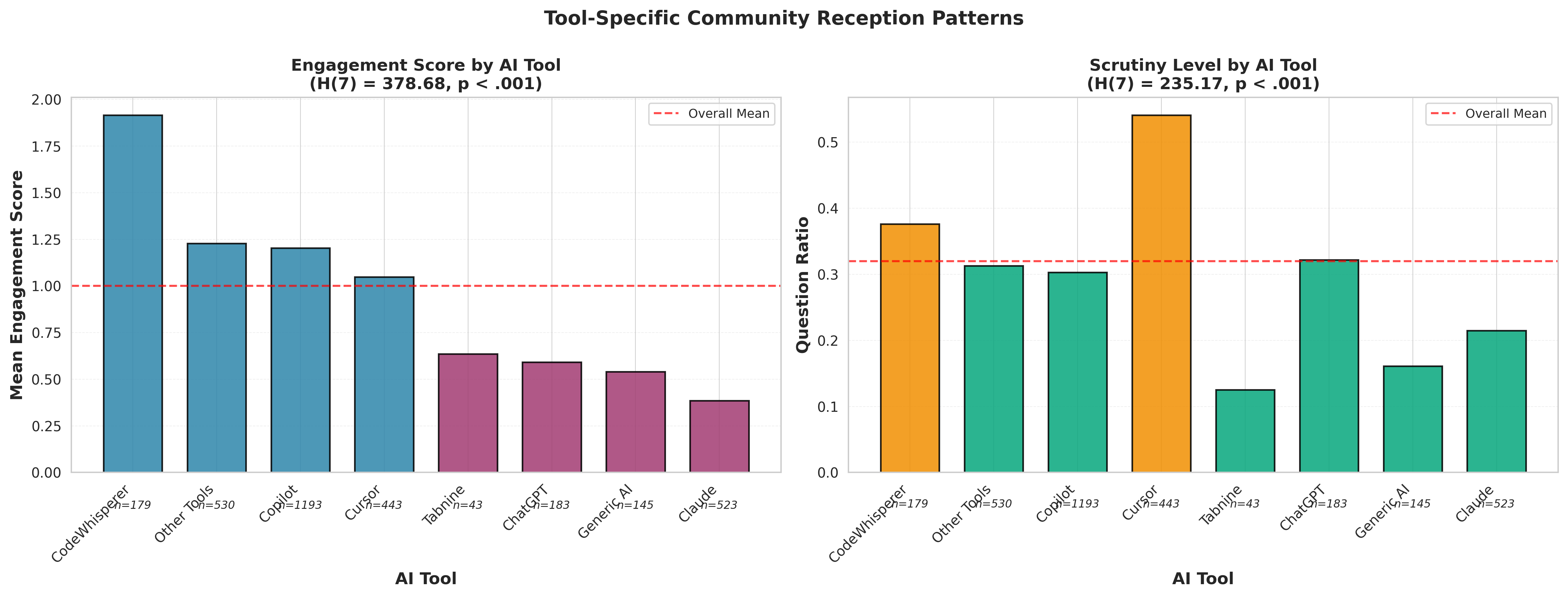}
\caption{Tool Reception Patterns}
\label{fig:tool_reception_patterns}
\end{figure}

These tool-specific patterns are illustrated in Figure \ref{fig:tool_reception_patterns}. CodeWhisperer-attributed contributions received the highest engagement ($M = 1.916$, $n = 179$), followed by Other Tools ($M = 1.228$, $n = 530$) and Copilot ($M = 1.204$, $n = 1{,}193$). Claude-attributed contributions received among the lowest engagement ($M = 0.386$, $n = 523$), despite Claude having the highest explicit attribution rate (72.3\%, from RQ1).

Cursor-attributed contributions exhibited the highest question ratio, with 54.1\% of comments containing questions ($M = 0.541$), compared to 30.3\% for Copilot and 21.5\% for Claude. Generic AI mentions and Tabnine showed the lowest question ratios (16.1\% and 12.5\% respectively). Sentiment scores were near-neutral across all tools (range: $-0.040$ to $-0.004$).

These findings answer RQ2 by demonstrating that attribution explicitness affects community reception through very small increases in questions (23\%) and comments (21\%), with small effect sizes. Tool choice appears to show far larger effects on reception than attribution type (20-30 times larger H-statistics), indicating that platform-specific norms overshadow attribution explicitness. Sentiment remains near-neutral across attribution types and tools, with no evidence of negative reception for explicit attribution.

\subsection{RQ3: Temporal Dynamics of AI Attribution}

The third research question examined the evolution of AI attribution practices and community responses from 2023 through 2025. Due to limited data in early 2023 ($n < 20$ per month), analyses focus primarily on trends from March 2024 onward when sample sizes became adequate for reliable temporal analysis ($n \geq 30$ per month).

Among AI-attributed commits with adequate monthly samples ($k = 25$ months), explicit attribution rates increased over time. Explicit disclosure increased from near-zero in early 2024 (0-2\% in Q1-Q2 2024) to 39.5\% by November 2025, with a positive temporal trend (Spearman's $\rho = .670$, $p < .001$). This increase was pronounced in late 2025, where explicit attribution jumped from 6.0\% in July to 58.9\% in September before stabilizing around 30-40\% in October-November. Overall, AI tool mentions remained relatively stable throughout 2024-2025 (80-100\% of commits per month).

Community reception showed limited temporal change (see Table \ref{tab:temporal_trends}). Only reaction counts showed significant temporal change, increasing over time ($\rho = .494$, $p = .005$). Other engagement metrics: comment counts ($\rho = .310$, $p = .095$), question ratios ($\rho = .340$, $p = .066$), and sentiment ($\rho = -.319$, $p = .086$), showed non-significant or marginally significant trends.

\begin{table}[htbp]
\caption{Temporal Trends in Community Reception (2023-2025)}
\label{tab:temporal_trends}
\centering
\small
\begin{tabular}{@{}p{5cm}rr>{\raggedright\arraybackslash}p{3cm}@{}}
\toprule
\textbf{Metric} & \textbf{Spearman's $\rho$} & \textbf{$p$-value} & \textbf{Direction} \\
\midrule
Engagement score & 0.154 & .417 & Decreasing \\
Reaction count & 0.494 & .005 & Increasing \\
Comment count & 0.310 & .095 & Decreasing \\
Sentiment & $-0.319$ & .086 & Increasing \\
Question ratio & 0.340 & .066 & Increasing \\
\bottomrule
\end{tabular}

\vspace{0.5em}
\begin{minipage}{\textwidth}
\centering
\footnotesize
\textit{Note.} Monthly aggregated data from 30 months spanning March 2023 through November 2025. Only reaction counts showed statistically significant temporal change.
\end{minipage}
\end{table}

\begin{figure}[htbp]
\centering
\includegraphics[width=0.9\textwidth]{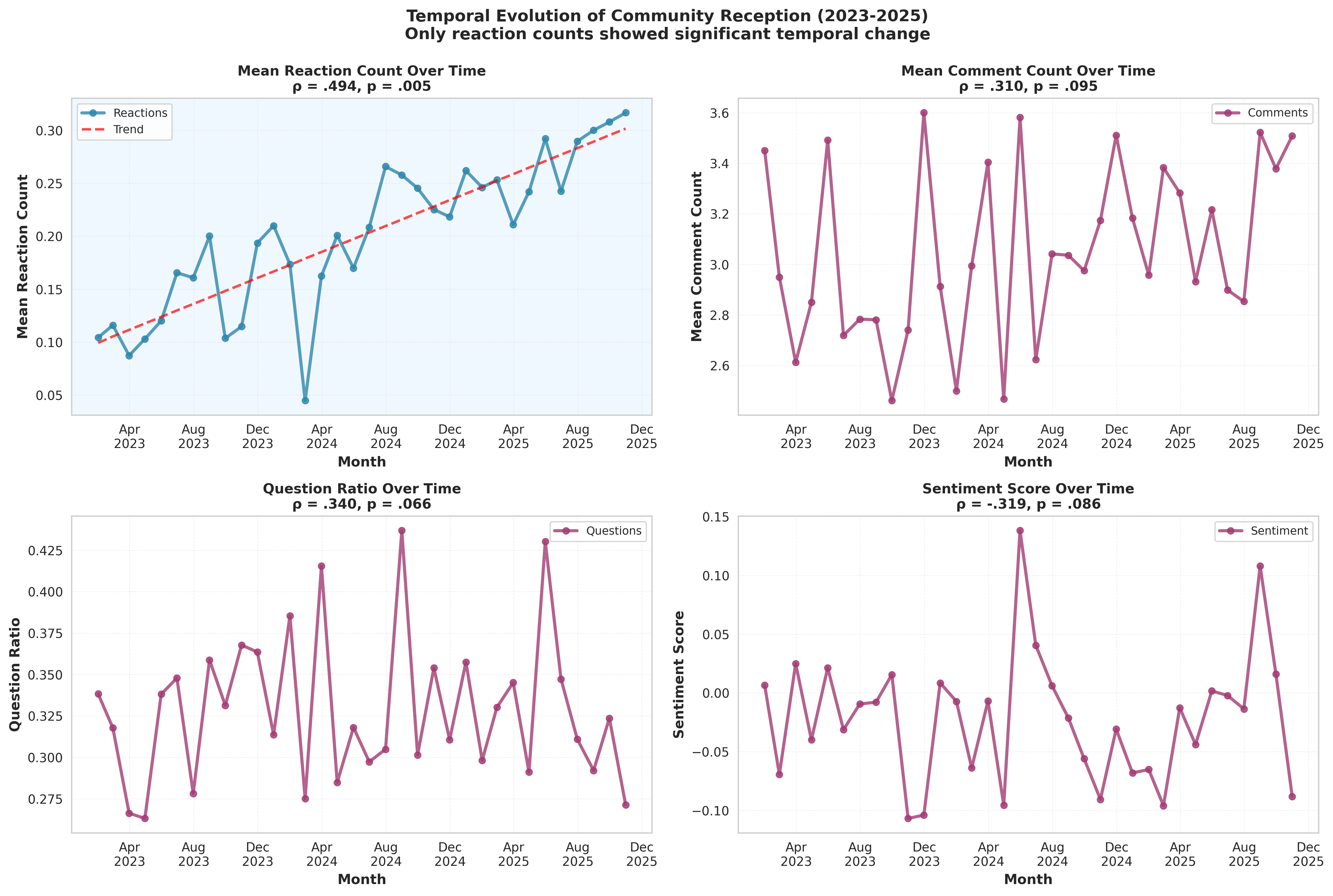}
\caption{Temporal Reception Trends}
\label{fig:temporal_reception}
\end{figure}

The temporal patterns in community reception metrics are illustrated in Figure \ref{fig:temporal_reception}. Attribution practices evolved differently across AI tools. Among tools with adequate temporal data, Copilot showed the strongest increase in explicit attribution over time ($\rho = .906$, $p < .001$), rising from 0\% in early 2024 to 7.4\% by late 2025. Claude maintained stable high explicit attribution throughout (60-77\%, $\rho = .100$, $p = .873$). ChatGPT showed minimal temporal change ($\rho = .300$, $p = .624$).

Attribution practices developed differently across AI tools. Among tools with adequate temporal data, Copilot showed the strongest increase in explicit attribution over time ($\rho = .906$, $p < .001$), rising from 0\% in early 2024 to 7.4\% by late 2025. Claude maintained stable high explicit attribution throughout (60-77\%, $\rho = .100$, $p = .873$). ChatGPT showed minimal temporal change ($\rho = .300$, $p = .624$). Similarly, attribution trends varied by repository age, with established repositories (created before 2023, $n = 1{,}787$) showing moderate increases (0\% to 20.2\%, $\rho = .589$, $p = .002$) and new repositories (created 2023-2025, $n = 12{,}283$) showing stronger increases (0\% to 40.4\%, $\rho = .703$, $p < .001$).

These observations answer RQ3 with evidence of dramatic increases in explicit attribution from 0\% to 40\% over the 2023-2025 period, with community reception remaining relatively stable. Tool-specific analyses showed differing pathways of explicit attribution, with Copilot showing strong temporal increases (0\% to 7.4\%) while Claude maintained stable high transparency (60-77\%). Repository age moderates adoption patterns, with newer repositories showing faster increases in explicit attribution than established repositories.

\section{Discussion}
This study focuses on how developers approach a central challenge in AI-assisted development, as they combine widespread AI use with strategic ambiguity about AI's role in writing code. Drawing on 14{,}300 GitHub commits, our analysis shows that AI tools are widely used (95\% of commits mention at least one), but explicit attribution remains limited: only 30\% of commits clearly credit AI with writing code. This ``attribution paradox'' does not look the same across tools; different AI assistants are associated with very different levels of explicit attribution. Claude users explicitly attribute AI in about 80\% of commits, whereas Copilot users do so in only 9\%, indicating sharply different tool cultures. Despite this selective transparency, developers who disclose AI involvement explicitly tend to receive constructive engagement rather than negative reactions. Overall, these patterns deepen our theoretical understanding of how AI tools are integrated into collaborative software workflows and offer practical guidance for managing attribution, communication, and trust in human--AI collaboration.

\subsection{The Attribution Paradox as Strategic Communication}
Our results show that fewer than one in three AI-attributed commits contain explicit disclosure of AI's role, suggesting that developers are engaging in impression management rather than straightforward transparency. This pattern aligns with what Goffman \cite{goffman1959presentation} describes as dramaturgical performance: developers selectively and carefully fine-tune what they disclose to shape how their work is seen, while still appearing authentic. Under this interpretation, the attribution paradox is not about developer dishonesty; it is about the complex social environments they navigate, where norms for disclosing AI involvement remain unsettled and can shift across projects, platforms, and audiences.

Tool-specific variations show that attribution norms differ by ecosystem instead of moving toward universal standards. The nine-fold difference between Claude and Copilot attribution rates suggests that communities anchored around each tool are building distinct norms, influenced by tool positioning, marketing narratives, and established usage patterns. Claude's high explicit attribution likely reflects Anthropic's focus on transparency and collaborative partnership, which has created community expectations about providing clear disclosure. In contrast, the integration of Copilot as an IDE extension may normalize AI suggestions as seamless developer workflow rather than external assistance requiring acknowledgment. These tool-specific cultures create challenges for developers who use multiple AI assistants, because each tool comes with its own expectations and attribution practices.

Explicit attribution is more common in well-known, established repositories, indicating that transparency expectations rise alongside project prominence and scrutiny. This pattern extends existing work on impression management \cite{goffman1959presentation} and transparency as social signaling in open-source development \cite{dabbish2012social,stewart2006impact}, where attribution practices have social functions beyond information dissemination to include being a form of reputation insurance and professionalism signaling. In such high-stakes contexts, explicit disclosure of AI assistance functions as a signal of accountability, even when developers have doubts about how the community will react to their use of AI.

Strategic attribution raises deeper questions of authorship and credit in collaborative software work, where developers continually try to balance transparency and efficiency. At the level of day-to-day practice, this tension shows up as competing pressures on developers: community norms that promote honest disclosure, worries that visible AI involvement might undermine how their code is perceived, the need to maintain speed without documenting every interaction, and ongoing uncertainty about what ``appropriate'' attribution should look like for different levels of support. Overall, this suggests that the attribution paradox is not a straightforward transparency problem, but an ongoing, situated negotiation of competing pressures, mediated through selective and context-dependent disclosure.

\subsection{Constructive Scrutiny Without Hostility}
Community responses challenge the idea that AI-generated code is being broadly rejected. Instead, we see a more hidden ``scrutiny tax'': when developers explicitly attribute code to AI, they receive a bit more follow-up in the form of questions and comments, even if the overall effect is small. This dynamic helps explain why attribution can feel paradoxical. Developers are deciding not only whether to reveal AI involvement, but also how clearly to do so, as they try to balance transparency with the desire to avoid extra scrutiny.

Three patterns challenge the narrative of avoidance. First, code reviews (the most time-consuming and consequential form of scrutiny) do not differ across attribution types, suggesting that the extra questions aimed at explicitly attributed commits are driven more by curiosity than by gatekeeping. Second, explicit attribution is linked to slightly more negative sentiment, but both groups remain close to neutral, suggesting that disclosure may create somewhat more critical discussion without tipping into hostility. Third, the choice of AI assistant plays a larger role than the clarity of attribution in shaping community responses, which underscores how strongly tool cultures and reputations structure interpretation.

At the tool level, the patterns we observe are even less straightforward. Claude, for example, shows high rates of explicit attribution but relatively low engagement, suggesting that disclosure by itself is not what drives greater scrutiny. In this case, explicit attribution appears to reassure rather than provoke, allowing developers to acknowledge AI involvement and reduce the need for further questioning. Seen this way, the attribution paradox looks different than a simple tension between disclosure and punishment. Developers may choose how explicit to be not simply to protect themselves from criticism, but because, in certain environments, clear attribution can limit the amount of further engagement required. What this reveals is the weight of tool ecosystems: developers work inside communities centered on particular platforms, and those platforms quietly set expectations about style and workflow that can have more impact than whether attribution is explicit or not.

In practice, this kind of constructive scrutiny does important social work. This social work happens through very specific questions---what the AI generated, how it was edited, and whether it should be treated differently from human-written code. The fact that we do not see strong negative sentiment directed at transparent developers challenges the assumption that openness will be used against them. In this regard, our results are consistent with arguments for greater algorithmic transparency \cite{burrell2016machine} and soften Ananny and Crawford's \cite{ananny2018seeing} concern that transparency mainly enables disciplinary action. Here, disclosure of AI involvement seems to promote accountability and trust instead of punishment, and developers who openly acknowledge AI use may be seen as more credible because they signal awareness of, and adherence to, evolving transparency norms.

\subsection{Rapid Norm Evolution Toward Transparency}
Across the three-year period, we see clear shifts in attribution norms, even though the attribution paradox still persists. During this period, explicit attribution shifted from almost zero in early 2024 to around 40\% by late 2025, with a sharp increase in mid-2025. This pattern looks like a tipping point in how developers talk about AI, likely driven by increased Claude use, more stable tooling, changing norms, and wider public discussion of transparency. Notably, this increase happened while overall AI tool mentions remained stable, which suggests that attribution is becoming more explicit rather than more common.

The relative stability in how communities respond suggests that disclosure habits and interpretive frameworks are evolving together. During the same period, reactions become more frequent and seem to function as a quick way to acknowledge contributions, while most other engagement measures remain fairly stable. This combination of steady reception and increasing explicitness means that, over time, clear attribution is becoming a normal part of practice rather than something exceptional. As a result, communities are converging on interpretive habits that treat disclosure as a cue for thoughtful engagement.

These temporal shifts align with a theory of norm formation through iterative communication \cite{postmes2000formation}. In the early stages, uncertainty about how to attribute AI creates a kind of sandbox in which developers can test different disclosure approaches. As developers watch how their peers attribute AI use and how communities react, shared expectations begin to form through collective sense-making rather than formal rules. As explicit attribution is received positively, it becomes easier to adopt, but the differences across tools show that norms are still forming within specific ecosystems rather than across the board.

As transparency becomes more common, it undermines the earlier view that strategic ambiguity would persist as the main tool for balancing conflicting concerns. Instead, the shift toward disclosure as a default seems to be driven by multiple reinforcing factors: tools that support attribution, community debates over best practices, visible leaders who model transparency, and encouraging responses that make openness feel safer. This gradual shift toward disclosure fits with the idea that online community practices are worked out through continuous collective negotiation \cite{butler2008community}.

Even as transparency grows overall, the lack of convergence across tools points to norms that are diverging instead of aligning, raising challenges for anyone hoping to define consistent practices across platforms. We see this fragmentation clearly in our tool-level trends: Copilot moves from almost no explicit attribution to around 7\%, whereas Claude starts and stays at a much higher level of transparency. A similar split appears at the project level, where newer repositories adopt explicit attribution more quickly than long-established ones. All of this suggests that attribution norms are not simply drifting toward a single standard, but instead have settled into tool- and project-specific patterns that continue to shape behavior. For developers using multiple AI assistants, these fragmented norms translate into an ongoing need to calibrate disclosure practices to the expectations of each tool ecosystem and project. Future research should examine whether these tool-specific norms are a temporary divergence or a more stable pattern rooted in how different tools position themselves and are used.

\subsection{Theoretical Implications}
These results illuminate how strategic communication operates within socio-technical systems. More specifically, they refine existing work on algorithmic transparency \cite{burrell2016machine,ananny2018seeing} and computer-mediated communication theory \cite{walther1996computer} by highlighting that disclosure is better understood as a strategic move that shapes interaction than as a neutral act of sharing information. Here, transparency is not only a feature of the platform, but also a decision developers make each time they describe how AI was involved. In text-based, asynchronous environments like commit messages, attribution is the point where expectations for openness meet developers' efforts to manage how their work is seen. This perspective widens the scope of CMC research from interpersonal relationship building to include how developers manage professional reputations and work out shared norms in collaborative settings. Future work on transparency should pay closer attention to the micro-dynamics of disclosure as communication (how people word, time, and frame attribution under competing pressures) rather than focusing only on whether information is technically accessible.

In addition to highlighting disclosure as a form of communication, our findings help explain how norms change even as technologies are moving rapidly. Across the three-year period we observe, we see a shift from uncertainty toward clearer expectations around transparency, suggesting that developers are actively collaborating to solve new coordination problems. From this perspective, socio-technical systems are settings where AI attribution practices are shaped more by distributed, stepwise changes than by formal policies or top-down rule-making.

\subsection{Practical Implications}
For developers, our results suggest that explicitly attributing AI assistance is often beneficial, even when it invites additional scrutiny. Extra questions may feel like added pressure, but they also open space for knowledge sharing and for signaling competence in managing AI tools responsibly. Rather than viewing disclosure as a reputational risk, developers can treat it as an opportunity to talk concretely about human-AI collaboration and to demonstrate professional accountability in how AI is used.

For platform designers, the findings argue for transparency by design rather than transparency by decree. Features such as attribution templates, standardized tags, or gentle prompts can make it easier to disclose AI assistance while preserving developer autonomy. At the same time, our evidence on constructive scrutiny suggests caution: blanket, mandatory disclosure rules could unintentionally flood developers with questions. Graduated mechanisms (where the level of required detail scales with the significance or visibility of a contribution) may offer a better balance between transparency, workload, and development speed.

Project maintainers and community leaders should promote explicit attribution as a beneficial norm while acknowledging the legitimate reasons for implicit disclosure in lower-stakes contributions. Code review enhances quality assurance, which is worth promoting. Moreover, the absence of negative sentiment indicates low social costs associated with AI attribution disclosure. Community guidelines can normalize this practice, and maintainers can model transparent practices to create cultures where disclosure is expected and valued.

Policymakers and governance bodies face challenges in implementing simple transparency mandates due to persistent strategic ambiguity despite transparency norms. This ambiguity suggests that disclosure cannot be fully regulated through policy alone. Instead, governance approaches should foster voluntary disclosure, support norm formation processes, and enable communities to adapt attribution practices to specific contexts.

\subsection{Limitations and Future Directions}
First, keyword-based detection limits the identification of commits where developers explicitly mention AI tools, potentially resulting in an underestimation of the total extent of AI assistance. However, our focus on attribution practices concerns visible signals rather than comprehensive usage measurement, which mitigates this limitation.

Second, causal direction between attribution type and community reception cannot be definitively established. While we theorize that attribution disclosure affects scrutiny, reverse causality remains possible. Temporal stability analyses and regression controls provide some causal purchase, but experimental manipulation would strengthen causal inference. Future research could use experimental vignettes systematically varying attribution explicitness to isolate causal impacts.

Third, the study period captures a specific historical moment of rapid AI tool adoption. Attribution norms may continue evolving, potentially converging toward higher transparency or fragmenting further by tool ecosystem. Longitudinal follow-up would track whether current transparency trends persist or plateau, and whether tool-specific differences narrow or widen. Major platform policy changes could disrupt observed patterns, making current findings temporally bounded.

Fourth, lexicon-based sentiment analysis may miss contextual nuance including sarcasm or cultural variation, though it remains adequate for large-scale classification. Future qualitative analysis could complement these patterns by revealing meanings communities attach to AI attribution.

Despite these limitations, our large-scale, multi-dimensional approach provides robust empirical foundation for understanding the AI attribution paradox. Effect consistency across tools, repository types, and time periods enhances confidence in findings and establishes baseline understanding for future research.

\section{Conclusion}

This study centers on a core paradox in AI-assisted coding: even as AI tools become routine, authorship remains strategically ambiguous. Using a dataset of 14,300 GitHub commits from the early AI adoption wave, we show that attribution practices operate as a crucial layer of strategic communication rather than as neutral record-keeping. These practices shape how transparency is performed, how collaboration unfolds, and how trust is negotiated in open-source development.

The data reveal three key empirical patterns. First, AI tools are referenced in nearly all commits in our sample (95\%), but explicit authorship attribution appears in only 30\% of them, with tool ecosystems diverging sharply: Claude commits are highly transparent (80\% explicit), whereas Copilot commits rarely are (9\%). Second, explicit attribution draws additional engagement (more comments and questions) but this ``scrutiny'' remains constructive, with no accompanying spike in negative sentiment or delay. Third, tracking these dynamics over time shows a rapid consolidation of norms around transparency, as communities gradually converge on shared ways of making sense of human-AI collaboration.

Theoretically, our findings reposition algorithmic transparency as an interactional practice: a way of communicating strategically about AI involvement rather than merely making information available. They also broaden computer-mediated communication theory by foregrounding how attribution practices shape professional standing and collaborative norms. The attribution paradox itself exposes underlying struggles over who gets credit, what counts as authorship, and how responsibility is distributed in human--AI work, highlighting selective disclosure as a key mechanism through which these tensions are managed.

On the practical side, our results suggests that developers can safely adopt explicit AI attribution as a routine practice, since it tends to spark constructive discussion rather than backlash. Platform designers should prioritize features that make voluntary disclosure easy and visible, instead of enforcing heavy-handed attribution rules. Community maintainers can frame explicit attribution as one element of responsible code stewardship, while still acknowledging that appropriate disclosure levels differ across tasks. For regulators and governance bodies, the findings underscore the value of supporting bottom-up norm development over imposing universal transparency mandates.

As AI coding assistants move from novelty to infrastructure, understanding how attribution works becomes increasingly important. The constructive scrutiny we observe suggests that communities are capable of integrating AI into their workflows while preserving quality and trust, and the rapid evolution of norms indicates that they can collectively solve new coordination problems under conditions of disruption. Future research should ask whether tool-specific cultures eventually converge on shared attribution practices or fragment further, whether current trends toward transparency persist, and how attribution mechanisms adapt as human-AI collaboration grows more complex.

Ultimately, the attribution paradox speaks to broader questions about work, credit, and collaboration with increasingly capable AI systems, questions that extend well beyond software development into many forms of knowledge work. Our findings suggest that communities are already learning to navigate these tensions, striking compromises between transparency and pragmatism, curiosity and efficiency, quality assurance and trust. This study captures an early moment in that transition, offering empirical grounding for understanding how human--AI collaboration is reshaping the social fabric of open-source development and providing concrete guidance for those tasked with steering these changes.

\bibliographystyle{unsrt}

\end{document}